\documentclass[sigconf]{acmart}
\AtBeginDocument{%
 \providecommand\BibTeX{{%
    \normalfont B\kern-0.5em{\scshape i\kern-0.25em b}\kern-0.8em\TeX}}}
\usepackage{enumitem}
\usepackage{graphicx}
\usepackage{epsfig}
\usepackage{epstopdf}
\usepackage{amssymb}
\usepackage{amsmath,amsthm}
\usepackage{comment}
\usepackage{longtable}
\usepackage{multirow}
\usepackage{booktabs}
\usepackage{tabularx}
\usepackage{algorithm2e}
\usepackage{makecell}
\usepackage{blindtext}
\usepackage{url}
\usepackage{listings}
\usepackage{subfig}

\usepackage{color}
\usepackage{textcomp}
\usepackage{gensymb}
\usepackage{array}
\usepackage{natbib}
\usepackage{colortbl}
\newcolumntype{P}[1]{>{\centering\arraybackslash}p{#1}}
\usepackage{caption}
\usepackage[flushleft]{threeparttable}
\definecolor{Gray}{gray}{0.9}
\usepackage{appendix}
\usepackage{stfloats}
\usepackage{float}
\newcommand{\modification}{\color{red}}
\newcommand{\modificationend}{\color{black}}

\definecolor{light-gray}{gray}{0.90}
 \makeatletter
\newcommand{\algorithmfootnote}[2][\footnotesize]{%
 \let\old@algocf@finish\@algocf@finish
 \def\@algocf@finish{\old@algocf@finish
    \leavevmode\rlap{\begin{minipage}{\linewidth}
    #1#2
    \end{minipage}}%
  }%
}
\makeatother
 
\usepackage[utf8]{inputenc}

\usepackage{todonotes}

\graphicspath{ {ACSAC/images/} }
\usepackage{wrapfig}
\usepackage{mathtools}
\newcommand{\Aegis}{{\textsc{\small{Aegis}}}\xspace}
\usepackage{tikz}

\iftrue
\newcommand{\leo}[1]{{\color{red}{\bf LB: }#1}}
\else
\newcommand{\leo}[1]{}
\fi

\makeatletter 
\newcommand\semiHuge{\@setfontsize\semiHuge{22.72}{27.38}}
\makeatother

\DeclareCaptionFont{black}{\color{black}}
\usepackage{textcomp}

\graphicspath{ {Images/} }

\setcopyright{acmcopyright}

\begin{document}
\title{\textsc{Aegis}: A Context-aware Security Framework\\ for Smart Home Systems}

\author{Amit Kumar Sikder, Leonardo Babun, Hidayet Aksu, and A. Selcuk Uluagac} 
    \affiliation{ 
      \institution{Cyber-physical System Security Lab \\ Florida International University}
    }
    \email{{asikd003, lbabu002, haksu, suluagac}@fiu.edu}

\begin{abstract}
Our everyday lives are expanding fast with the introduction of new Smart Home Systems (SHSs). Today, a myriad of SHS devices and applications are widely available to users and have already started to re-define our modern lives. Smart home users utilize the apps to control and automate such devices. Users can develop their own apps or easily download and install them from vendor-specific app markets. App-based SHSs offer many tangible benefits to our lives, but also unfold diverse security risks. Several attacks have already been reported for SHSs. However, current security solutions consider smart home devices and apps individually to detect malicious actions rather than the context of the SHS as a whole. The existing mechanisms cannot capture user activities and sensor-device-user interactions in a holistic fashion. To address these issues, in this paper, we introduce \Aegis, a novel context-aware security framework to detect malicious behavior in a SHS. Specifically, \Aegis observes the states of the connected smart home entities (sensors and devices) for different user activities and usage patterns in a SHS and builds a contextual model to differentiate between malicious and benign behavior. We evaluated the efficacy and performance of \Aegis in multiple smart home settings (i.e., single bedroom, double bedroom, duplex) with real-life users performing day-to-day activities and real SHS devices. We also measured the performance of \Aegis against five different malicious behaviors. Our detailed evaluation shows that \Aegis can detect malicious behavior in SHS with high accuracy (over 95\%) and secure the SHS regardless of the smart home layout, device configuration, installed apps, and enforced user policies. Finally, \Aegis achieves minimum overhead in detecting malicious behavior in SHS, ensuring easy deployability in real-life smart environments. 
\end{abstract}

\copyrightyear{2019} 
\acmYear{2019} 
\acmConference[ACSAC '19]{2019 Annual Computer Security Applications Conference}{December 9--13, 2019}{San Juan, PR, USA}
\acmBooktitle{2019 Annual Computer Security Applications Conference (ACSAC '19), December 9--13, 2019, San Juan, PR, USA}
\acmPrice{15.00}
\acmDOI{10.1145/3359789.3359840}
\acmISBN{978-1-4503-7628-0/19/12}


\maketitle
\thispagestyle{fancy}
\section{Introduction}

The capabilities of the smart home devices have evolved from merely controlling lights and opening garage doors to connecting our living spaces to the cyber world~\cite{stojkoska2017review, lee2014securing, chitnis2016investigative}. Such functionality provides more autonomous, efficient, and convenient daily operations~\cite{fernandes2017security}. For instance, sensor-activated lights offer energy efficiency while smart locks and motion-activated cameras offer a more secure home environment. Compared to early Smart Home Systems (SHS) with fixed device setup procedures and limited functionalities, modern SHSs have adopted a more user-centric, app-based model. Similar to the smartphone ecosystem, SHS's users can download apps from the vendor's app market and easily set up and control the smart devices, which makes SHSs more popular and versatile than ever~\cite{smarthome1}. 

\par The integration of programming platforms with smart home devices surely enhances the functionalities of SHSs, but it also exposes the vulnerabilities of the devices to the attackers. Attackers can release malicious apps in third-party markets and public repositories (e.g., GitHub) easily. Then, careless users can download and utilize them for their devices. From here, the attackers can exploit smart home devices in several ways: they can perform denial-of-service attacks to obstruct normal operations of SHS~\cite{ransom}, they can compromise one device in SHS and get access to other connected devices~\cite{fernandes2016security}, they can even leak personal information such as unlock code of a smart lock and gain physical access to the home~\cite{jia2017contexiot, saint-taint-analysis}. Recently, a repository of malicious apps in different smart home platforms has been published exhibiting several vulnerabilities of the current smart home app development ecosystem~\cite{iotbench}. Nonetheless, a security solution that detects these emerging threats associated with SHSs does not exist and is direly needed. 

Recent studies have proposed the implementation of enhanced permission models for SHSs, which depends on specific user permission~\cite{jia2017contexiot} or the analysis the source code of the apps to detect vulnerabilities, which is only effective against specific types of attacks~\cite{saint-taint-analysis}. Moreover, existing solutions focus on the detection of malicious activities that affect smart home devices and apps individually. However, a more holistic approach that also considers user activity contexts and sensor-device-user interactions (e.g., movement directions, sensors activated, rooms involved) is needed. For example, if a user walks from the bedroom to the hallway, s/he may activates multiple devices and sensors along his/her path (i.e., walking context) in a specific sequence: moving towards the bedroom door, opening the door, entering the hallway, closing the door, and reaching the hallway. A user cannot simply skip any of these steps and reach the hallway directly from the bedroom. Again, device actions in a SHS are correlated with each other which can be observed from a context-aware model. For example, a smart light triggered by the motion sensor can be verified by checking the user’s presence in the home using a presence sensor. A contextual awareness of devices and applications that considers these types of sensor-device-user interactions can provide valuable information about malicious activities occurring in the SHSs, something that is missing in current smart home solutions.

\par To address these emerging threats and shortcomings of SHSs, we present \Aegis, a novel context-aware security framework to detect malicious behavior in a SHS. \Aegis observes the changing patterns of the conditions (active/inactive) of smart home entities (sensors and devices) for different user activities and builds a contextual model to detect malicious activities. Here, context-awareness refers to the ability of \Aegis to understand the changes in sensors and devices' states due to ongoing user activities and determine if the behavior in the SHS is benign or not. Smart home devices are typically configured with different sensors to provide autonomous control and uninterrupted operations. Thus, different sensors in a SHS can sense user activities (motion, opening doors, etc.) and trigger associated devices to perform pre-defined tasks. \Aegis correlates this sensor-device relation with different user activities and builds a context-aware model to define benign user behavior. \Aegis also uses app context to understand the trigger-action scenarios between smart home entities (sensors and devices) and automatically upgrades the framework if new devices are added to the SHS. As a security framework, \Aegis observes the current states (active or inactive) of smart home sensors and devices and checks with the learned user behavior to detect any malicious behavior. Specifically, \Aegis utilizes a Markov Chain-based machine learning technique to detect malicious behavior. Additionally, \Aegis uses an action management system to alert the users in the event of malicious behavior and considers user responses to improve the context-aware model for better accuracy (adaptive training mode). We tested \Aegis in real SHSs scenarios where 15 different users performed typical daily activities in three different home layouts generating over 55000 sensor-device correlated events. Furthermore, we considered different device settings (sensor-device relations), apps, and user policies to evaluate the performance of \Aegis against five different threats. Our extensive evaluation demonstrates that \Aegis can detect different threats to SHS with high accuracy and F-score (over 95\%). In addition, \Aegis achieves minimum overhead in terms of latency and resource usage making \Aegis compatible for real-life deployment. \par

\noindent\textit{\textbf{Contributions:}} Our main contributions are noted 
as follows: 
\begin{itemize}[wide = 0pt]
    \item \textbf{\Aegis.} We present a novel context-aware security framework to detect malicious activity 
    in SHS. We capture sensor-device co-dependence in smart home to understand the context of the user activity and detect malicious behavior. Additionally, we implemented an action management system to alert users about \Aegis's findings. 
    \item \textbf{User-specific configurations.} We designed \Aegis to support different smart home layouts and configurations. \Aegis allows easy integration of new devices and apps creating app contexts and reconfiguring the training data automatically. We also introduced an adaptive training model to improve the detection mechanism from user responses automatically. 
    \item \textbf{High accuracy and minimal overhead.} Through a detailed evaluation, we demonstrated how \Aegis can detect different malicious activities in a SHS. Our results show that \Aegis can achieve high accuracy and F-score and impose minimum overhead in the system. 
\end{itemize}
\textit{\textbf{Organization:}} The rest of the paper is organized as follows: In Section~\ref{background}, we present the background information. Then, we discuss the adversary model in Section~\ref{adv}. Section \ref{frame} details \Aegis's architecture and Section~\ref{eval} evaluates the efficacy of \Aegis in detecting different malicious behavior in SHS. In Section~\ref{discussion} and Section~\ref{related}, we discuss how different types of users will be benefited by deploying \Aegis in real-life SHS and the related work, respectively.  Finally, Section~\ref{conclusion} concludes the paper.

\section{Background}\label{background}

\begin{figure}[t!]
  \centering
    \includegraphics[width=0.45\textwidth]{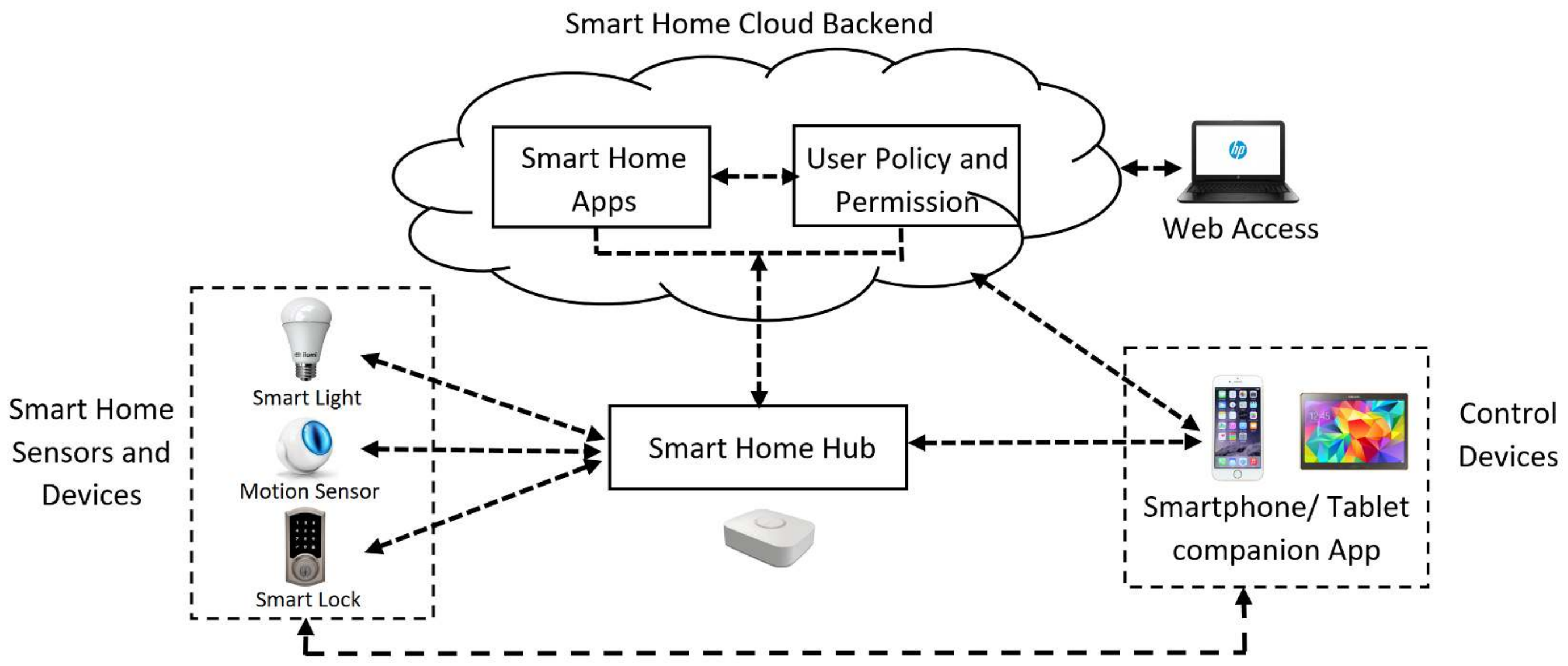}
      \caption{A smart home environment and its major components.}
      \label{smarthome}
      \vspace{-0.2in}
\end{figure}

In this section, we describe the components of the SHS that we assume for \Aegis. We also detail different features used in \Aegis to detect different malicious activities in SHS. In Figure~\ref{smarthome}, a typical architecture of a SHS is shown. A SHS has four basic building blocks as shown in Figure~\ref{smarthome}. The first block of the SHS comprises sensors and devices in the system. These smart home devices and sensors are connected to each other via a smart hub. As there is no generic interoperability standard among smart home devices, the hub provides a common access point for all the entities in the SHS. The hub is connected to both cloud backend service and smartphone/tablet companion app. Users can use the smartphone app to control the smart home entities or install different apps from the app stores. Indeed, we can group SHS architectures in two main categories: a cloud-based architecture where the installed apps run in the cloud backend (e.g., SmartThings), and hub-based architecture where the installed apps run the hub locally (e.g., Apple HomeKit). 

\subsection{New Design Features Considered by \textsc{Aegis}}

\begin{figure}[t!]
  \centering
  \vspace{-0.2cm}
    \includegraphics[width=9cm, height = 4.5cm]{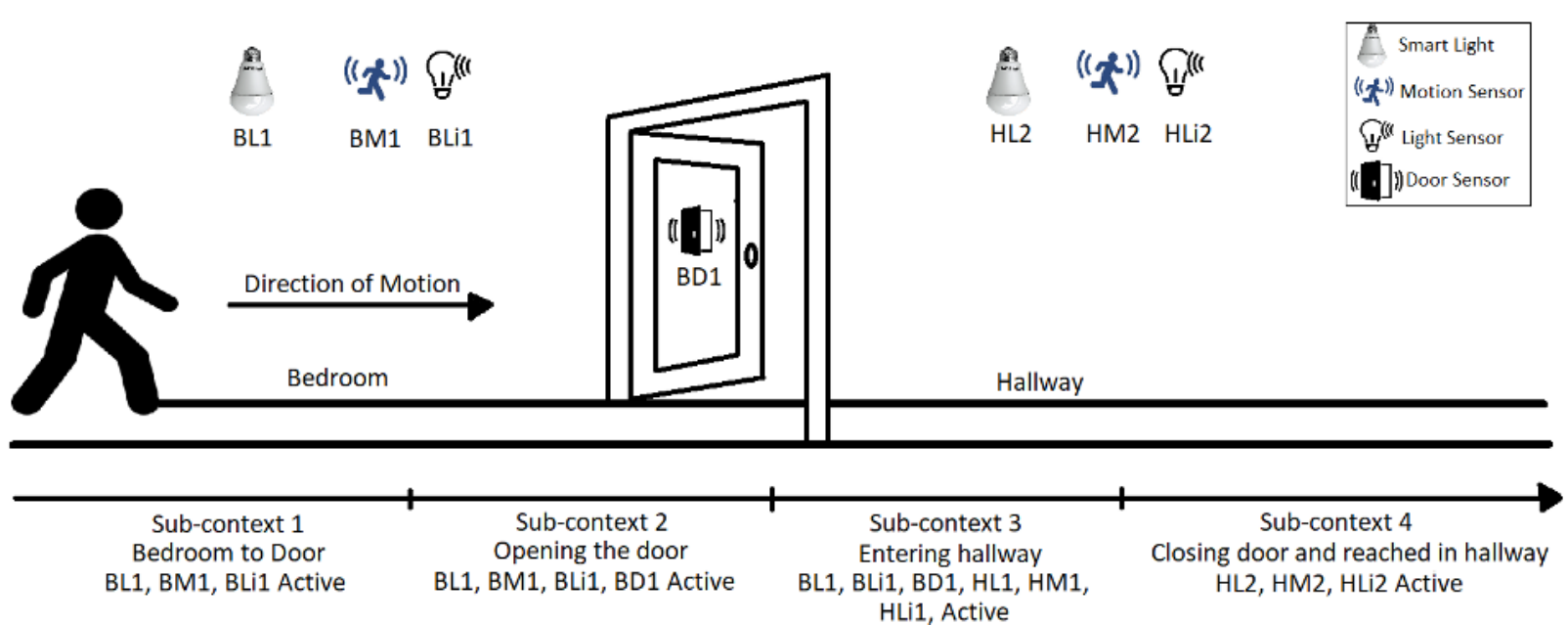}
    \vspace{-0.5cm}
      \caption{Context-awareness feature, which is not considered in existing SHS.}
      \label{context}
      \vspace{-0.4cm}
\end{figure}

\noindent\textbf{Context-awareness.} Context-awareness refers to the ability of a system to use situational and environmental information about the user, location, and devices to adapt its operation accordingly~\cite{203854, jia2017contexiot, sikder2019context3}. In a SHS, all the sensors and devices follow different trigger-action scenarios to perform tasks. Here, sensors are used to provide input in the devices (trigger) and devices take autonomous decisions (actions) based on these inputs. When a user performs a task in a SHS, several smart home sensors and devices may become active in a sequential pattern. The pattern of active devices and sensors is different but specific for distinct user activities. Existing SHS cannot observe these patterns in sensors' and devices' states over time and can not understand the context of the user activity. For example, while a user moves from one bedroom to a hallway, several devices and sensors become active in a sequential manner (Figure~\ref{context}): moving towards bedroom door (sub-context 1: BL1, BLi1, BM1 are active), bedroom door opens (sub-context 2: BL1, BLi1, BM1, BD1 are active), entering the hallway (sub-context 3: BL1, BLi1, BD1, HLi2, HL2, HM2 are active), bedroom door and light close and reaches the hallway (sub-context 4: HLi2, HL2, HM2 are active). To complete the activity (moving from the bedroom to the hallway), the user must follow the sub-contexts in the same sequential pattern. The user cannot skip one specific sub-context and move to the next one to complete the activity. For instance, the transition from sub-context 1 to sub-context 4 is not possible as a user cannot go to the hallway from the bedroom without opening the door. Motivated by this, \Aegis is designed to understand this property of SHS to build a context-aware model for different user activities and usage patterns and differentiates between benign and malicious activities of smart home devices and sensors. 

\noindent\textbf{Sensor-device co-dependence} In a SHS, sensors, and devices can be configured as independent entities. However, they work in a co-dependent manner to provide autonomous functionalities. For instance, smart lights can be configured with motion sensors to light up when motion is sensed. Here, the smart light depends on the input from the motion sensor while the motion sensor alone cannot provide any significant function in a SHS. The functions of devices and sensors create a co-dependent relationship with each other. In this way, sensors and devices in the SHS can build many-to-many co-dependent relationships. However, existing SHSs do not consider this co-dependent relationship and can not visualize the context of a user activity by observing the usage pattern of smart home entities. In short, sensors and devices in a SHS are configured as independent components, but in reality, they are function-wise co-dependent. \Aegis considers these relations to build the context of the user activities in a SHS.

\noindent\textbf{User activity-device correlation.} In a SHS, different users utilize and control smart home devices in multiple ways. For instance, a user can set a security camera to take pictures whenever a motion is detected in the associated motion sensors. On the other hand, users control devices in multiple ways. For example, a user can unlock a door by using the smartphone app or entering the code manually. Here, the state of the lock can be determined by user activity on the smartphone or by using a presence sensor to detect the user near the smart lock. In short, by observing the user activities in a SHS, it is possible to determine the normal operation of smart home devices. One can define normal or malicious user behavior with the user activity-device correlation. Current SHS cannot correlate user activity and device actions correctly, which is considered as a feature in \Aegis to differentiate benign and malicious activities.

\section{Problem Scope and Threat Model}\label{adv}
We introduce the problem scope and articulate the threat model.

\subsection{Problem Scope}
We assume a fully automated SHS with several smart home devices and sensors. Here, the following sensor-device triggering rules are configured - the smart lights are configured with motion sensors, the smart smoke detector is configured with smoke sensor. The SHS allows manual device control by the users (e.g., unlocking smart lock with PIN). We also assume that the user utilizes customized third-party apps to control the devices. Furthermore, the SHS has more than one user authorized to control the devices in the system. We assume the following incidents happening throughout the day in the SHS - (1) one user is walking inside the home but the lights are not triggered by the motion sensors, (2) one user is trying to unlock the smart lock using PIN code, (3) a fire alarm is being triggered in the system, (4) a smart light inside the house executes a blinking pattern.

We propose \Aegis as a novel security framework that builds a context-aware model based on user activities to determine benign and malicious incidents in the SHS. \Aegis answers several questions that may arise from the above-mentioned incidents - (1) What is the reason for no activity in the smart light?, (2) Is an attacker is trying to unlock the door using PIN code?, (3) Is the fire alarm being triggered by a malicious app?, (4) What caused the smart light to blink and what is the intent of this activity? \Aegis differentiates between normal and malicious activities happening in a SHS. Furthermore, \Aegis detects malicious activities occurring in a device by observing the ongoing activities of all the connected devices in the SHS. 

\subsection{Threat Model}
\Aegis considers \textit{anomalous user behaviors} (e.g., unauthorized users changing the device states) that may disrupt the normal functionality of the SHS. Also, device vulnerabilities that may cause device malfunction or open doors to threats like impersonation attacks and false data injection attacks are considered by \Aegis. Additionally, this work assumes carelessly-designed and malicious smart home applications that may cause unauthorized or malicious activities in the SHS. These malicious activities may facilitate side channel and denial-of-service (DoS) attacks. In Appendix~\ref{threattable}, we present specific examples of attack scenarios that are used later to evaluate the effectiveness of \Aegis (Section \ref{eval}).

\begin{table}[t!]
\centering
\footnotesize
\fontsize{6}{8}\selectfont
\resizebox{0.45\textwidth}{!}{
\begin{tabular}{ccp{3.5cm}}
\toprule
\textbf{\begin{tabular}[c]{@{}c@{}}Threat\end{tabular}} & \textbf{\begin{tabular}[c]{@{}c@{}}Attack Method\end{tabular}} & \textbf{\begin{tabular}[c]{@{}c@{}}Attack Example\end{tabular}}\\ \hline
\midrule
\multirow{3}{*}{Threat-1}    & \multirow{3}{*}{\begin{tabular}[c]{@{}c@{}}Impersonation \\attack\end{tabular}}   & An unauthorized user steals the unlock code of a smart lock and try to unlock the door. \\ \hline
\multirow{4}{*}{Threat-2}   & \multirow{4}{*}{\begin{tabular}[c]{@{}c@{}}False data \\injection\end{tabular}}  & A malicious smart home app can exist in the SHS and inject forged data to perform malicious activities.\\ \hline
\multirow{4}{*}{Threat-3}   & \multirow{4}{*}{\begin{tabular}[c]{@{}c@{}}Side channel \\attack\end{tabular}}   & A malicious smart home app can exist in the SHS and perform legitimate, yet vulnerable side-channel activities (switching on when no one is around in vacation mode) which can be harnessed by other apps (considered malicious) in the system or the attacker himself.\\ \hline
\multirow{4}{*}{Threat-4}   & \multirow{4}{*}{\begin{tabular}[c]{@{}c@{}}Denial-of-service \end{tabular}} & A malicious smart home app installed in the SHS and impede normal behavior of the smart home devices.\\ \hline
\multirow{4}{*}{Threat-5}   & \multirow{4}{*}{\begin{tabular}[c]{@{}c@{}}Triggering a malicious \\app\end{tabular}}  & A malicious smart home app can exist in the system which can be triggered by a specific activity pattern or device action (e.g., switching a smart light in a specific on/off pattern) in a smart home environment.\\
\bottomrule
\end{tabular}}
\caption{Summary of the threat model considered for \Aegis.}
\label{adversary}
\vspace{-0.8cm}
\end{table}

The information leakage caused by a compromised device or untrusted communication channel in the SHS are considered out of scope of \Aegis. We also assume that the data collected from the devices and central management system (e.g., Hub, cloud, etc.) is not compromised. In Table~\ref{adversary}, we summarize the threat model used later in \Aegis's performance evaluation (Section~\ref{eval}).

\begin{figure*}[t!]
  \centering
    \includegraphics[width=0.85\textwidth]{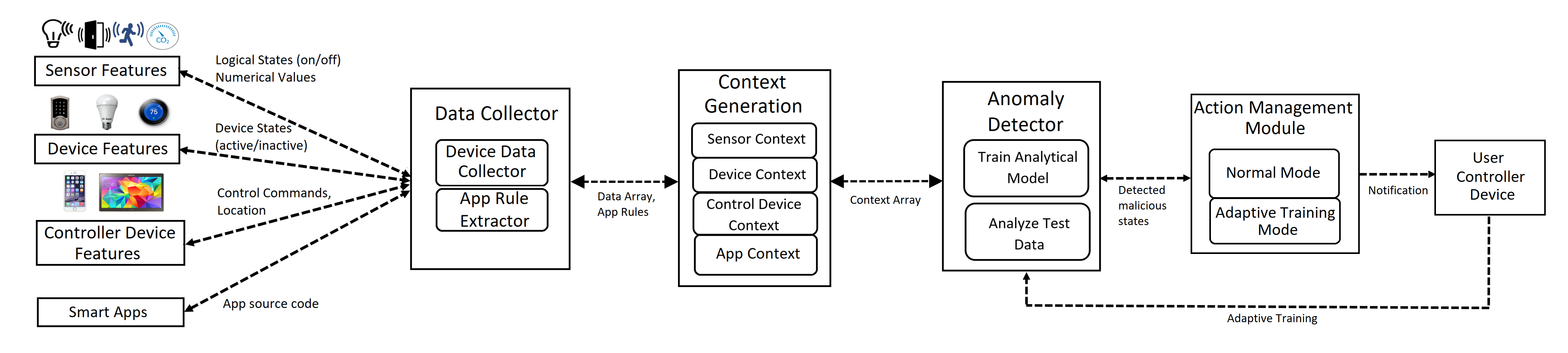}

  \caption{The architecture of \Aegis.}
      \label{overview}
\end{figure*}

\section{\textsc{Aegis} Framework}\label{frame}

\Aegis has four main modules: (1) data collector, (2) context generator, (3) data analysis, and (4) action management (Figure~\ref{overview}). First, \Aegis collects data from smart home entities (sensors and devices) for day-to-day user activities. The \textit{data collector module} uses an app to collect all the device states (active/inactive) from the hub. Additionally, this module collects rules generated by different smart apps using the \textit{app rule extractor}. 

The device state data is used to understand the context of the user activities and feeds the \textit{context generation module}. This module creates context arrays depending on usage patterns and the predetermined user policies in the smart apps. Each context array contains overall information of the user activities and device states in the SHS.

The context arrays generated in the context generation module are used by the \textit{anomaly detector module} to implement machine learning-based analysis and build the context-aware model of the SHS. Additionally, \textit{anomaly detector module} decides whether or not malicious activities occur in the SHS. 

Finally, the malicious activities detected by the anomaly detector module are forwarded to the \textit{action management module}. This module notifies the users regarding the unauthorized activities. Also, it offers adaptive training mode where users can validate any false positive or false negative occurrence and re-train the detection model to improve the performance of \Aegis. 

\subsection{Data Collector Module}
The data collector module has two sub-modules: device data collector and app data collector. 
\subsubsection{Device Data Collector}
\Aegis collects data from smart home devices and sensors using the data collector module. In a SHS, there can be multiple devices and sensors connected through a hub and operating in a co-dependent manner. Data collector of \Aegis collects the state of these devices (active or inactive) autonomously and forwards these data to the context generation module. Based on the type of data, the collected data is governed by:
\begin{equation}
    Data\ array,\ E = \{S, D, M\},
\end{equation}
where S is the set of features extracted from the sensors, D is the set of features extracted from the devices, and M is the features extracted from the associated controller devices (e.g., smartphone, smart tablet) in a SHS. We describe the characteristics of these features below.
\begin{itemize}[wide = 0pt]
    \item \textit{Features extracted from sensors (S):} An SHS can comprise several sensors such as motion and light sensors. They sense changes in the vicinity of the devices and work as input to multiple devices. Sensor data can be both logical states (e.g., motion sensor) and numerical values (light sensor). For \Aegis, we consider both logical states and numerical values of sensors to create the context of user activities.
    \item \textit{Features extracted from devices (D):} In a SHS, several devices can be connected with each other and also with different sensors. These devices remain active based on user activities in a smart home environment. \Aegis observes the daily activities of users and collects the device state data (active/inactive state) to build the context of the associated activity. 
    \item \textit{Features extracted from controller devices (M):} In a SHS, Smartphone or tablet works as a control device to the SHS and users can control any device using the associated smart app of the smart home. \Aegis considers any control command given from the controller device as a feature to understand the context of user activity. Additionally, the location of the connected controller device can also work as an input to control multiple devices. For example, a thermostat can be configured to the desired temperature whenever the smartphone of the user is connected to the smart home network. \Aegis considers the location of the controller device as a feature to build the context of user activities.
\end{itemize}

As user activities on a SHS can vary based on the number of users, \Aegis considers multi-user settings to understand the user activity contexts correctly. Moreover, user activities also change based on the daily routine of users. For this, in the data collection process, \Aegis also offers time-based activity settings (weekday and weekend settings).  \par
\vspace{-0.1in}
\subsubsection{App Rule Extractor}\label{conf}
Modern SHSs offer an app-centric model where users can install different apps to automate the functions of smart home devices. These apps mostly define a trigger-action scenario for specific devices. For instance, an app can automate a smart light by configuring it with a motion sensor or light sensor. Here, the sensors work as a trigger and the state of the smart light (on/off) refers to the action. These trigger-action scenarios can represent the app context which can be used to validate the user activity context in a SHS. Additionally, the app context is also used to train the analytical model for new devices in SHS. \Aegis uses a static analysis tool, logic extractor, to extract the app logic which is used to determine the app context in a SHS. Logic extractor takes the source code of an app as the input and extracts the trigger-action scenarios enforced in the app as the output. For example, if a smart light is configured with a contact sensor, \Aegis extracts the following logic from the app (sample app is given in Appendix~\ref{appcontext}. 
\begin{lstlisting}[caption= trigger-action Scenario of a sample app,label=listing-logic]
Trigger: Contact1
Action: Switch1
Logic 1: contact1 = on, light1 = on
Logic 2: contact1 = off, light1 = off
\end{lstlisting}

\subsection{Context Generation Module}
The data collector module forwards the collected data to the context generation module to build the context of different user activities. Then, the context generation module maps and aggregates the data to build context arrays. Each context array consists of information on the usage patterns in the SHS for different activities, which can be used for further analysis and to determine malicious activities in the system. The context array modeling process has the following steps:
\begin{itemize}[wide = 0pt]
\item \textit{Context of sensors:} Sensor features collected in the data collector consists of both logic state (on/off) and numerical values. \Aegis observes the sensor data and generates the conditions of the sensors. These conditions represent the changing pattern of the sensor. If the current sensor value is different than the previous one, \Aegis considers this as an active condition that is represented as 1. Similarly, conditions labeled as inactive are represented as 0. 
\item \textit{Context of devices:} Data collector of \Aegis collects device state (active/inactive) data for every connected device. These device state data are converted to logical states (1 represents active and 0 represents inactive) to build the context of user activities in a SHS.
\item \textit{Context of controller devices:} There are two features of the controller device (e.g., smartphone, tablet, etc.) that are collected by \Aegis: Control command for the devices and the location of the controller device. For any command from the smartphone/tablet, \Aegis considers the active condition of smartphone/tablet which is represented as 1 in context array or 0 otherwise. An SHS allows two different states to represent the location of the controller device - \textit{home} and \textit{away}. Home location indicates that the controller device is connected to the home network and away represents otherwise. \Aegis represents the "home" location of the smartphone as 1 and the "away" location as 0 in the context array.
\end{itemize}
The final context array can be represented as follows:
\begin{equation}\label{contextarray}
\footnotesize
    Context\ Array,\ C = [\{S_1, S_2, ..., S_X\}, \{D_1, D_2, ..., D_Y\}, \{M_1, M_2\}],\\
\end{equation}
where $S_1, S_2, ..., S_X$ captures the conditions of X number of sensors in the SHS, $D_1, D_2, ..., D_Y$ the conditions of Y number of sensors in the SHS, and $M_1, M_2$ the conditions of smartphone/tablet in the SHS.\par

Context generation module also generates the app's context. As most of the app's logic represents a trigger-action scenario, the context generation module converts the logic in a binary representation. For example, the logic extracted from the app presented in Listing 1 is given below:
\begin{lstlisting}[caption=Generated app context of a sample app,label=listing-context]
App Context 1: contact1 = 1 , Light1 = 1
App context 2: contact1 = 0, Light1 = 0
\end{lstlisting}
Here, for the contact sensor, 1 and 0 represent the contact state from "open" or "close" respectively. Similarly, for the smart light, 1 and 0 represent the light state from "on" or "off", respectively. These app contexts are used to validate the sensor-device co-dependence captured in the context array. Additionally, these app contexts are used to update the training dataset whenever a new device is added to the SHS.

\subsection{Anomaly Detector Module}
In the third module, \Aegis takes context arrays generated in the context generation module as input and trains a Markov Chain-based machine learning model which is used to detect malicious activities in SHS. \par
 
\begin{figure}[t!]
  \centering
    \includegraphics[width=0.4\textwidth]{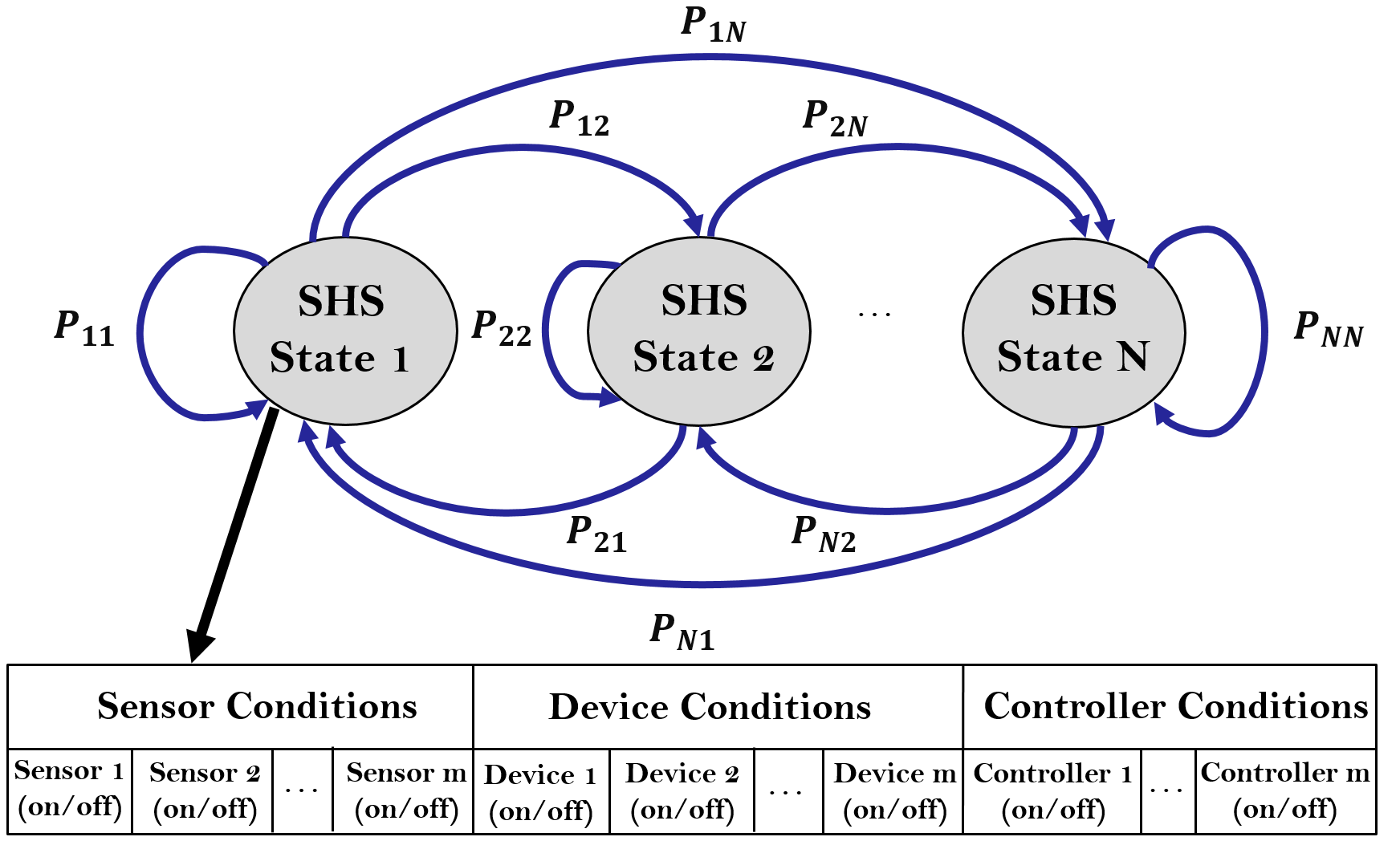}
      \caption{Markov Chain model for \Aegis.}
      \vspace{-0.4cm}
      \label{markov}
\end{figure}

The Markov Chain model is based on two main assumptions: (1) the probability of occurring a state at time $t+1$ only depends on the state at time $t$ only and (2) the transition between two consecutive states is independent of time. \Aegis uses this Markov Chain model to illustrate a series of events in a SHS. Here, a series of events denotes user activity and usage pattern and the state represents the context array at a specific time $t$ generated in the context generation module. The probabilistic condition of Markov Chain model is shown in Equation \ref{probabilityMarkov}, where $X_t$ denotes the state at time $t$ for a user activity in the SHS~\cite{keilson2012markov}.
\begin{equation}\label{probabilityMarkov}
\footnotesize
\begin{split}
\begin{aligned}
P(X_{t+1} = x| X_1 = x_1, X_2 = x_2 ..., X_t = x_t) = P(X_{t+1} = x| X_t = x_t) , \\ 
when,\ P(X_1 = x_1, X_2 = x_2 ..., X_t = x_t) > 0
\end{aligned}
\end{split}
\end{equation}
\Aegis considers the context array given in Equation \ref{contextarray} as an array of variables and observes its changes over time. For every user activity on a SHS, several context arrays are created. These arrays follow a different but specific pattern for different user activities. Each element of the context array represents the condition of a smart home entity (active/inactive states of sensor, device, or smartphone). For a distinct time, $t$, we consider the combination of all the smart home devices' and sensors' condition as binary output (1 for the active state of an entity and 0 for the inactive state). Thus, the number of total states (A) will be the exponent of 2 and can be represented as a n-bit binary number, where n is the total number of entities in the SHS. Let assume $P_{ij}$ denotes the transition probability of the system from state $i$ at time $t$ to state $j$ at time $t+1$. If the SHS has $n$ number of entities and $m=2^n$ states in the system, the transition matrix of the Markov Chain model can be illustrated by Figure~\ref{markov}. Here, each transition probability from one state to another state represents an element of the transition matrix.

If the SHS has \textit{${X_0, X_1, \ldots, X_T}$} states at a given time \textit{${t= 0, 1, \ldots, T}$}, the elements of the transition matrix can be shown as ${P_{ij}} = \frac{N_{ij}}{N_i}$ where $N_{ij}$ denotes the number of transition from $X_t$ to $X_{t+1}$, $X_t$ is the state at time $t$, and $X_{t+1}$ is the state at time $t+1$~\cite{ye2000markov}. Instead of predicting the next state using the Markov Chain model, \Aegis determines the probability of transition between two states in the SHS at a given time. We train the Markov Chain model with the generated context arrays from the context generation module and construct the transition matrix. Using this transition matrix, \Aegis determines the probability of transition from one state (i.e., context array) to another state over time. For example, in Figure~\ref{context}, the transition between sub-context 1 and sub-context 2 is valid as the user can perform this activity. However, the transition from sub-context 1 to sub-context 4 is invalid as the user cannot go from the bedroom to the hallway without going through sub-context 2 and 3. Thus, \Aegis defines benign or malicious device behavior based on user activities.

\subsection{Action Management Module}\label{action}
Finally, the action management module notifies the users in the event of malicious activity in the SHS. Action management module has two operation modes - detection mode and adaptive training mode.
\begin{itemize}[wide = 0pt]
    \item \textit{Detection mode:} In the detection mode, \Aegis pushes a notification in the controller device (smartphone, smart tablet) to notify the users if malicious activity is detected. \Aegis provides the device ID and the installed app names to the user for further action.
    \item \textit{Adaptive training mode:} As \Aegis builds a contextual model from user activities, it is important to verify the correct context of an ongoing user activity~\cite{perera2014context}. In a SHS, users can perform different activities in an irregular pattern. For example, a guest may come to the house which will introduce some new activity patterns in the SHS. These abrupt data patterns may cause a higher false positive rate in the contextual model. To address this issue, \Aegis offers the adaptive training mode, a user feedback process to improve the performance. In the adaptive training mode, whenever a malicious activity is detected, \Aegis sends a notification to the controller device (smartphone, smart tablet, etc.) for user confirmation. Users can either confirm the malicious activity from the controller device or mark the activity as benign. If the user confirms the activity as benign, \Aegis label that activity context and train the framework automatically. Hence, \Aegis can correctly and automatically improve the training dataset by adding irregular or new user activity pattern.
\end{itemize}

\vspace{-0.1cm}
\section{Performance Evaluation}\label{eval}

In the evaluation of \Aegis, we consider several research questions:
\begin{itemize}[wide=0pt, nolistsep]
    \item[\textbf{RQ1}] What is the performance of \Aegis in different smart home layouts and devices? (Sec~\ref{layouts})
    \item[\textbf{RQ2}] What is the impact of different apps, policies, and configurations on the performance of \Aegis? (Sec.~\ref{configs})
    \item[\textbf{RQ3}] What is the impact of different user behavior on the performance of \Aegis? (Sec.~\ref{behavior})
    \item[\textbf{RQ4}] What are the performance overhead introduced by \Aegis in a SHS? (Sec.~\ref{over})
\end{itemize}

\subsection{Evaluation Setup and Methodology}

We obtained appropriate Institutional Review Board (IRB) approval to collect daily usage data of a SHS with multiple users. We implemented a test setting where the users had the freedom to design their own smart home environment and perform regular daily activities in a timely order. While collecting the user activity data, we considered five features to enrich our dataset: \textit{Anonymous User ID}, \textit{User Role}, \textit{Smart Home Layout}, \textit{Activity Day-time}, and \textit{User Policy}. For each user, we assigned an anonymous ID to ensure privacy. We also assigned a specific user role to each participant to understand the context of the user activities in a multi-user scenario. As there exist different home layouts in real scenarios, we considered three different home layouts (single bedroom apartment, two-bedroom home, and duplex home) and let the users choose their layout and smart home devices. Moreover, user activities in a SHS depend on the user's daily routine which may change for different days of the week. We considered this and collected the user activities performed in both weekdays and weekends, separately. Finally, as current smart home platforms let users define multiple policies to control devices, we allowed the users to define their own policies in the SHS during the data collection process.

\begin{table*}[t!]
\vspace{-0.2cm}
\centering
\fontsize{6}{8}\selectfont
\resizebox{\textwidth}{!}{
\begin{threeparttable}
\begin{tabular}{cccccccccccccc}
& \multicolumn{6}{c}{\textbf{Normal Training}} & \multicolumn{6}{c}{\textbf{Adaptive Training}}\\ \hline
\multicolumn{1}{|c|}{\textbf{Smart Home Layout}} & \textbf{\begin{tabular}[c]{@{}c@{}}Recall\end{tabular}} & \textbf{\begin{tabular}[c]{@{}c@{}}FN\end{tabular}} & \textbf{\begin{tabular}[c]{@{}c@{}}TN\end{tabular}}   & \textbf{\begin{tabular}[c]{@{}c@{}}FP\end{tabular}} & \textbf{Accuracy} & \textbf{F-score} & \multicolumn{1}{|c}{\textbf{Recall}} & \textbf{\begin{tabular}[c]{@{}c@{}}FN\end{tabular}} & \textbf{\begin{tabular}[c]{@{}c@{}}TN\end{tabular}}   & \textbf{\begin{tabular}[c]{@{}c@{}}FP\end{tabular}} & \textbf{Accuracy} & \multicolumn{1}{c|}{\textbf{F-score}} \\ \hline

Single Bedroom Home   &  0.95  &  0.05  &  1  &  0      &  0.9547 &  0.9604 & 0.97 & 0.03 & 1 & 0 & 0.9712  & 0.9847\\ 
Double Bedroom Home   &  0.93  &  0.07  &  1  &  0      &  0.9340 &  0.9655 & 0.96 & 0.04 & 1 & 0 & 0.964  & 0.9795 \\ 
Duplex Home   &  0.91  & 0.09   & 1   &  0      & 0.9119  & 0.9529  & 0.96 & 0.04 & 1 & 0 & 0.9614  & 0.9688\\ 
\hline
\end{tabular}
\end{threeparttable}}
\caption{Performance evaluation of \Aegis in different smart home layouts.}
\label{Layout}
\vspace{-0.4cm}
\end{table*}

We chose Samsung SmartThings platform to create the smart home environment because of its large app market and compatibility with other smart devices~\cite{smartthingbusiness}. We considered the most common devices in our SHS. A detailed list of devices that were used in our experiments is given in Appendix~\ref{devices}. We collected data from 15 different individuals with different user roles, user policies, and smart home layouts. Our dataset consisted of over 55000 events collected in a 10-day period. We implemented a custom app as part of \Aegis's data acquisition module that uses the \textit{ListEvent} command from SmartThings API to collect the device logs. For collecting app context data from installed apps, we used a static analysis tool available online \cite{saint-taint-analysis}. We created an app context database which consists of 150 official \textit{Samsung SmartThings} apps. Whenever users install an app in SHS, \Aegis searches for existing app context in the database and adds the context into the training dataset for data validation purpose. For any third-party app, users can manually use the source code of the app in \Aegis and generate the app context which is later added in the database.\par

\begin{table*}[t!]
\centering
\vspace{-0.3cm}
\fontsize{6}{8}\selectfont
\resizebox{\textwidth}{!}{
\begin{threeparttable}
\begin{tabular}{cccccccccccccc}
& \multicolumn{6}{c}{\textbf{Normal Training}} & \multicolumn{6}{c}{\textbf{Adaptive Training}}\\ \hline
\multicolumn{1}{|c|}{\textbf{Smart Home Layout}} & \textbf{\begin{tabular}[c]{@{}c@{}}No of\\Controllers\end{tabular}} & \textbf{\begin{tabular}[c]{@{}c@{}}Recall\end{tabular}} & \textbf{\begin{tabular}[c]{@{}c@{}}FN\end{tabular}} & \textbf{\begin{tabular}[c]{@{}c@{}}Precision\end{tabular}}   & \textbf{\begin{tabular}[c]{@{}c@{}}FP\end{tabular}} & \textbf{Accuracy} & \textbf{F-score} & \multicolumn{1}{|c}{\textbf{Recall}} & \textbf{\begin{tabular}[c]{@{}c@{}}FN\end{tabular}} & \textbf{\begin{tabular}[c]{@{}c@{}}Precision\end{tabular}}   & \textbf{\begin{tabular}[c]{@{}c@{}}FP\end{tabular}} & \textbf{Accuracy} & \multicolumn{1}{c|}{\textbf{F-score}} \\ \hline

\multirow{3}{*}{\begin{tabular}[c]{@{}l@{}}Single Bedroom\\Home\end{tabular}} & 2    & 0.9472      & 0.0528      & 1         & 0     & 0.9477   & 0.9729 & 0.9685 & 0.0315 & 1 & 0 & 0.9711  & 0.9839\\ \cline{2-14}
& 3    & 0.9399      & 0.0601      & 1         & 0     & 0.9405   & 0.9690 & 0.9564 & 0.0436 & 1 & 0 & 0.96  & 0.9777\\ \cline{2-14}
& 4    & 0.9041      & 0.0959      & 0.96         & 0.04     & 0.9352   & 0.9312 & 0.9482 & 0.0588 & 1 & 0 & 0.9525  & 0.9734\\ \hline
\multirow{3}{*}{\begin{tabular}[c]{@{}l@{}}Double Bedroom\\Home\end{tabular}} & 2    & 0.9222      & 0.0778      & 1         & 0     & 0.9229   & 0.9595 & 0.9654 & 0.0346 & 1 & 0 & 0.9682  & 0.9823\\ \cline{2-14}
& 3    & 0.9058      & 0.0942      & 0.9529         & 0.0471     & 0.9062   & 0.9288 & 0.9523 & 0.0477 & 0.9785 & 0.0215 & 0.9545  & 0.9652\\ \cline{2-14}
& 4    & 0.8806      & 0.1194      & 0.8941         & 0.1059     & 0.8807   & 0.8873 & 0.9476 & 0.0524 & 0.96 & 0.04 & 0.9486  & 0.9537\\ \hline
\multirow{3}{*}{\begin{tabular}[c]{@{}l@{}}Duplex Home\end{tabular}} & 2    & 0.9017      & 0.0983      & 1         & 0     & 0.9038   & 0.9483 & 0.958 & 0.042 & 1 & 0 & 0.9615  & 0.9785\\ \cline{2-14}
& 3    & 0.8901      & 0.1099      & 0.9238        & 0.0762    & 0.8909   & 0.9067 & 0.9512 & 0.0488 & 0.975 & 0.025 & 0.9531  & 0.9629\\ \cline{2-14}
& 4    & 0.8694      & 0.1306     & 0.8857         & 0.1143     & 0.8698   & 0.8775 & 0.9388 & 0.0612 & 0.953 & 0.047 & 0.94  & 0.9458\\ \hline
\hline
\end{tabular}
\end{threeparttable}}
\caption{Performance evaluation of \Aegis in different multi-user scenarios.}
\label{multi-user}
\vspace{-20pt}
\end{table*}

For collecting the malicious dataset, we created six different attack scenarios, and their associated smart home apps based on the adversary model presented in Section~\ref{adv} (more details in Appendix \ref{threattable}). Additionally, we added some malfunctioning devices (e.g., a smart lock without power, fused smart light, etc.) in the SHS to test \Aegis in cases that include device malfunction. We collected 24 different datasets (4 dataset for each attack scenario) for a total of over 10000 events. We used 75\% of the benign user data to train the Markov Chain model. Then the remaining 25\% of data along with the malicious dataset was used in the testing phase which is a common practice~\cite{sikder2019context, GoogleML, Amazon}. Finally, to evaluate \Aegis, we utilized six different performance metrics: True Positive Rate or recall rate (TPR), False Negative Rate (FNR), True Negative Rate or precision rate (TNR), False Positive Rate (FPR), Accuracy, and F-score. Details of these performance metrics are given in Appendix~\ref{metrics}.

\vspace{-0.1in}
\subsection{Evaluation with Different Home Layouts}\label{layouts}
To evaluate \Aegis in different smart home layouts, we consider two important criteria (1) different smart home layouts, (2) multiple numbers of users. A SHS can have different smart home layouts and different number of devices. We tested the efficiency of \Aegis in a multi-user environment and different smart home layouts.\par
\noindent\textbf{Different smart home layouts:} User activities in a smart home can vary depending on the home layout as different layouts can lead to different usage patterns. To evaluate \Aegis, we considered three different layouts: single bedroom home, double bedroom home, and duplex home. Here, we considered a single authorized smart home user in different layouts. We collected data from 15 different users in these layouts. Table~\ref{Layout} presents the evaluation results associated with different smart home layouts. We can observe that accuracy and F-score for different layouts vary from 96-91\% and 97-95\%, respectively. \Aegis also achieves high TP (96-91\%) and TN rate (100\%) irrespective of layouts. One can safely confirm that variation in different layouts has a minimal impact on the performance of \Aegis. Table~\ref{Layout} also shows how the performance of \Aegis improves in adaptive training mode. Here, whenever the controller device (e.g., smartphone, tablet, etc.) is connected in the smart home network, we infer the user is in home location and use adaptive training mode. One can notice that the accuracy of \Aegis increases from 95\% to 97\% in adaptive training mode for single bedroom layout. For double bedroom and duplex home, \Aegis achieves 96.4\% and 96.1\% accuracy respectively. As adaptive training mode uses user validation to reduce FP and FN events, F-Score increases to approximately 97\% for all three layouts. In summary, \Aegis can achieve accuracy and F-score over 95\% for all three smart home layouts. \par

\begin{figure*}[t!]
\vspace{-0.2cm}
\centering
\subfloat[True Positive (TP)]{\includegraphics[width=0.22\textwidth]{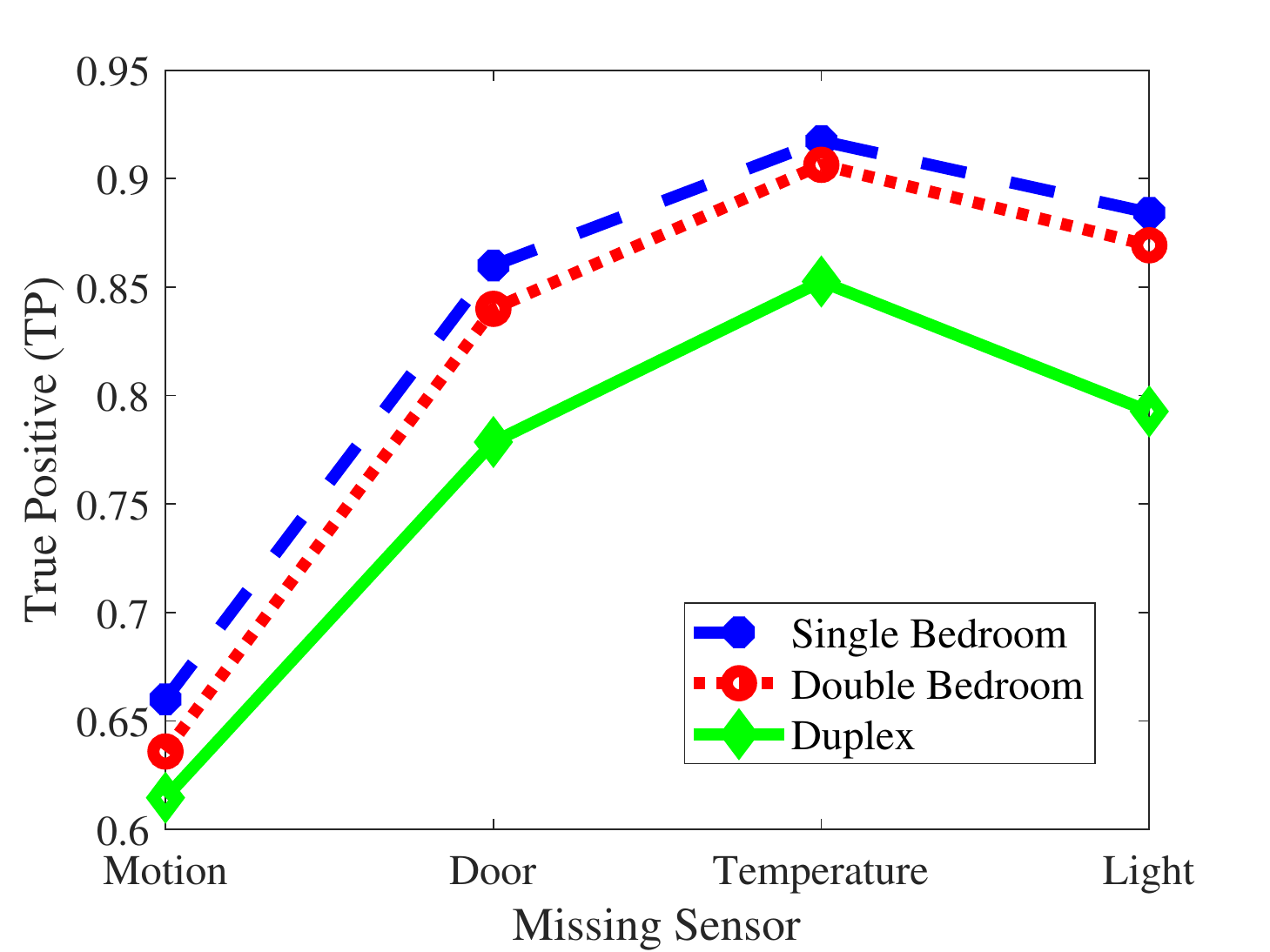}\label{fig:tp}}
\subfloat[False Negative (FN)]{\includegraphics[width=0.22\textwidth]{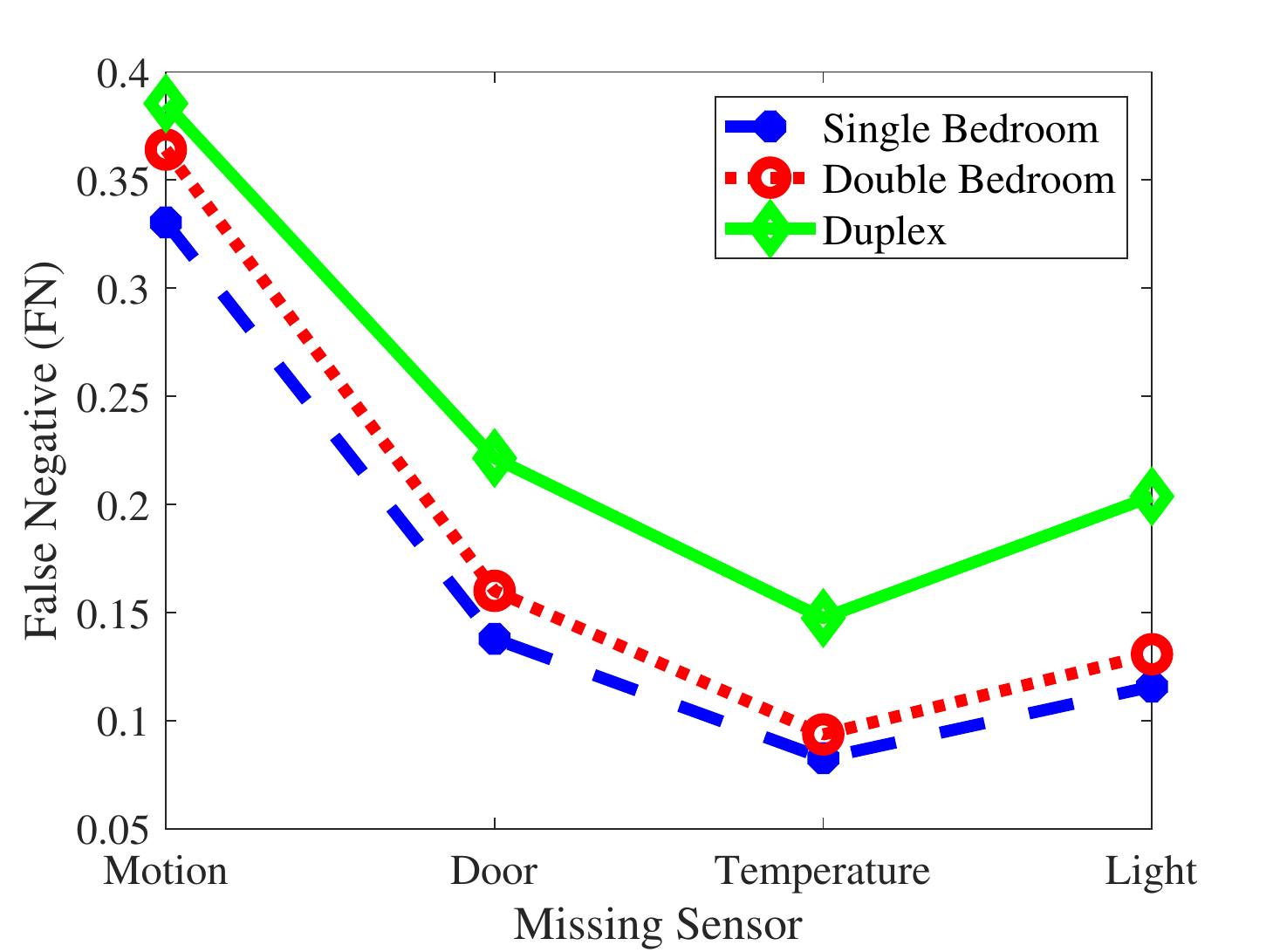}\label{fig:fn}}
\subfloat[Accuracy]{\includegraphics[width=0.22\textwidth]{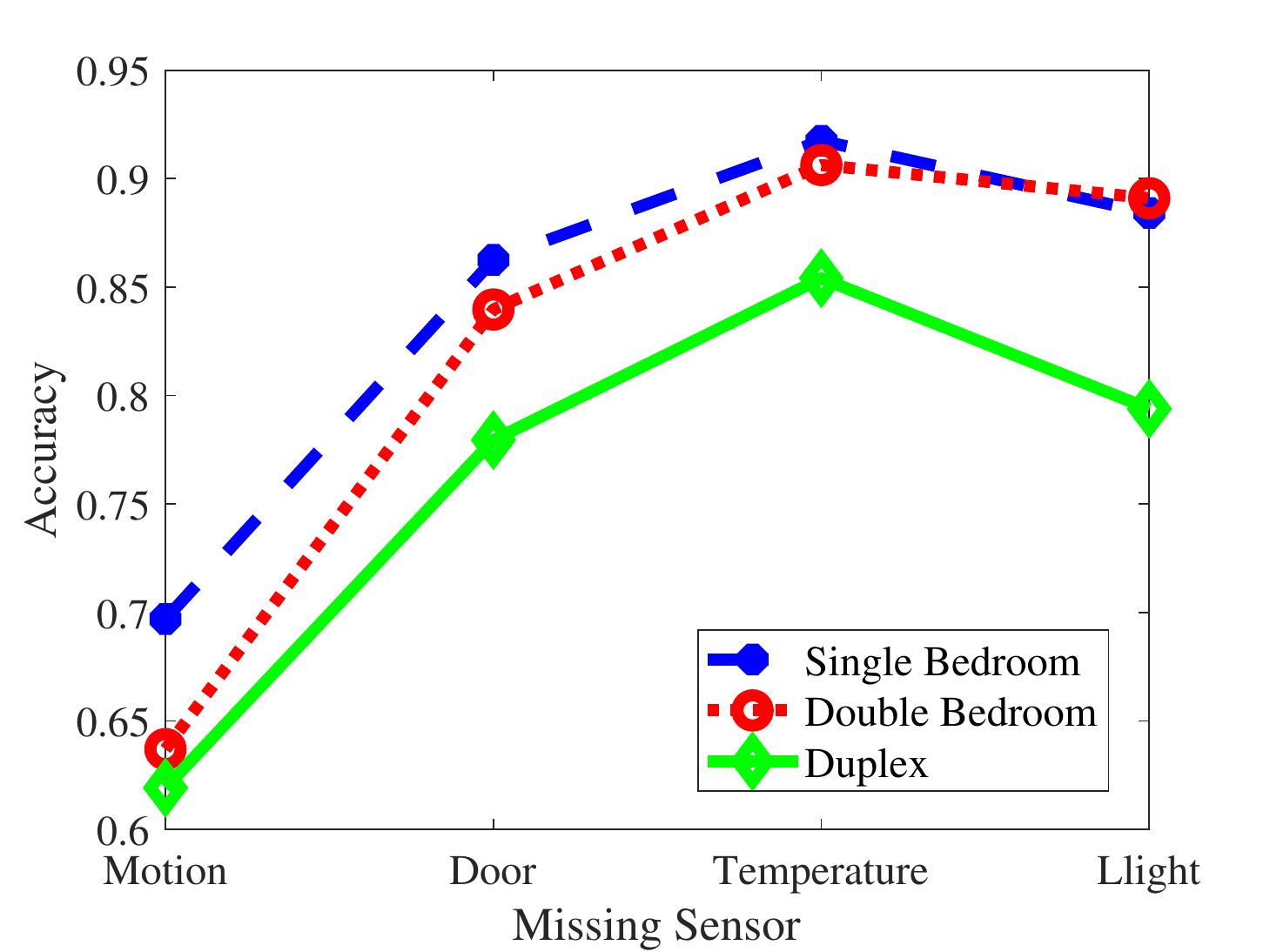}\label{fig:acc}}
\subfloat[F-Score]{\includegraphics[width=0.22\textwidth]{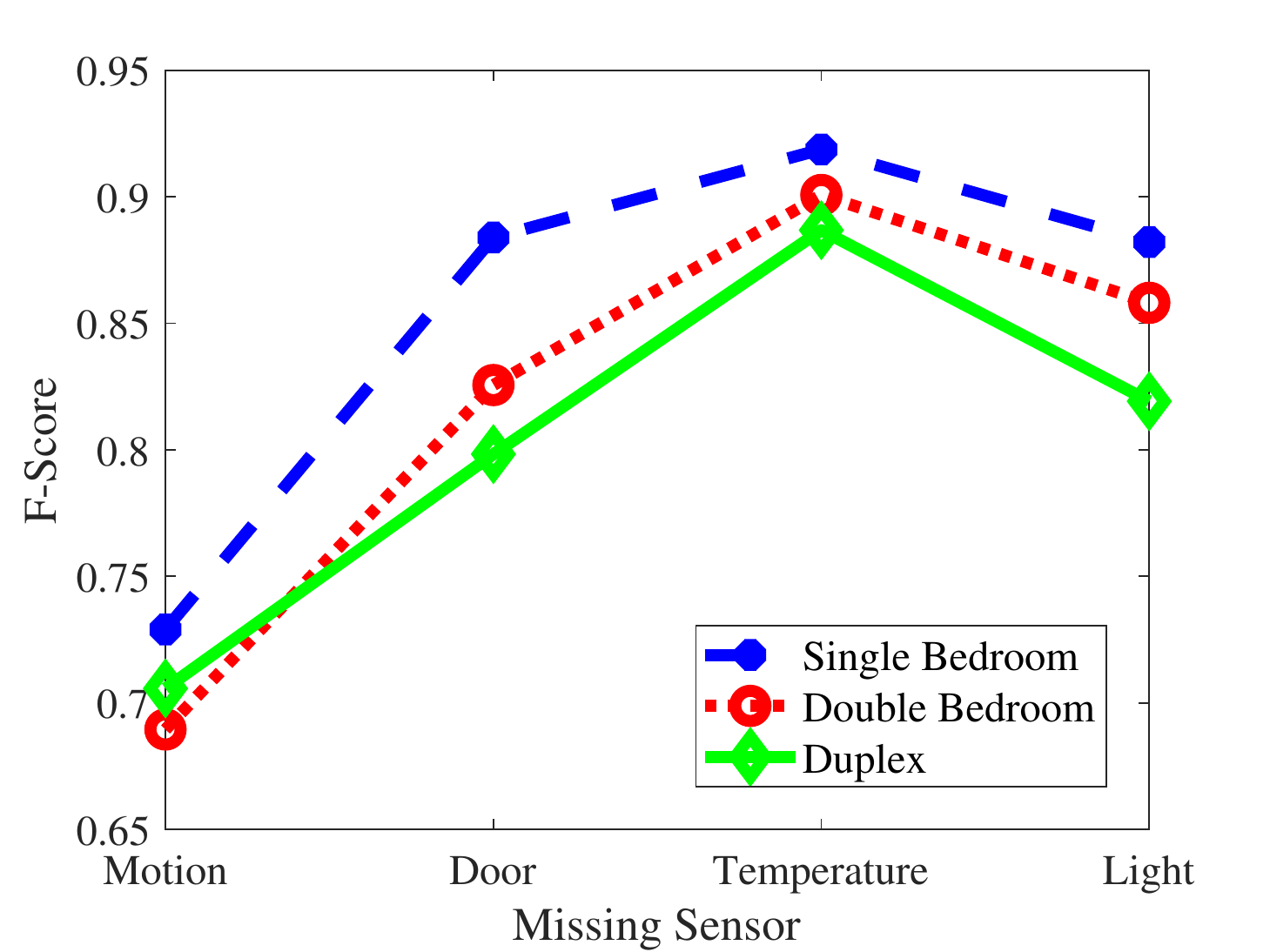}\label{fig:fs}}
\vspace{-0.2cm}
\caption{Performance evaluation of \Aegis with different sensors.}
      \label{sensor}
      \vspace{-0.3cm}
\end{figure*}

\noindent\textbf{Different number of authorized users:} Smart home platforms allow users to add more than one authorized user for the same SHS. Hence, a SHS can have multi-user scenarios with different user activities happening at the same time. To evaluate this setting of the smart home in \Aegis, we collected data from several multi-user settings with different users performing their daily activities at once. We used different smart home layouts with several multi-user scenarios (two, three, and four authorized controllers/conflicting users) in our data collection process. Additionally, we performed the aforementioned attack scenarios to collect malicious dataset and tested the efficiency of \Aegis in different multi-user environments. Table~\ref{multi-user} illustrates the detailed evaluation of \Aegis in different smart home settings. For single bedroom layout, we can observe that accuracy and F-score reach the peak (0.9477 and 0.9729, respectively) for the two users setup. If we increase the number of authorized users in the SHS, the accuracy gradually decreases with an increasing FN rate. Similarly, for double bedrooms and duplex home layout, \Aegis achieves the highest accuracy and F-score for two authorized users' setup. Both accuracy and F-score decreases while the FN rate increases as the number of authorized users increases. The highest accuracy achieved in two bedrooms and duplex home layouts are 92.29\% and 90.38\%, respectively. As different users interact with smart home devices in varied ways, the FN rate increases with the number of users in the system. To minimize the number of FN events, we implement the adaptive training mode in \Aegis. In a multi-user scenario, a notification is pushed in all the controller devices if \Aegis detects a malicious event in adaptive training mode. All the authorized users can confirm the event based on their activities and \Aegis trains the analytical model with validated data. One can notice from Table~\ref{multi-user} that \Aegis achieves the highest accuracy and F-Score (97\% and 98\%, respectively) for two users setup in single bedroom layout. Adaptive training mode also decreases FN rate approximately by 38.6\% and increases the accuracy to 96\% and 95.25\% for three and four authorized user scenarios respectively. For two bedroom and duplex home layout, adaptive training mode also increases the efficiency of \Aegis. Adaptive training mode reduces FN and FP rate approximately by 60\% while accuracy and F-Score increases to approximately 96\% and 98\% respectively in a double bedroom and duplex home layout. In summary, \Aegis can minimize the effect of conflicting user activities in a multi-user scenario in adaptive training mode while increasing efficiency.  

\begin{figure*}[t!]
\vspace{-0.1in}
  \centering
\subfloat[Single Bedroom Layout]{\includegraphics[width=0.20\textwidth]{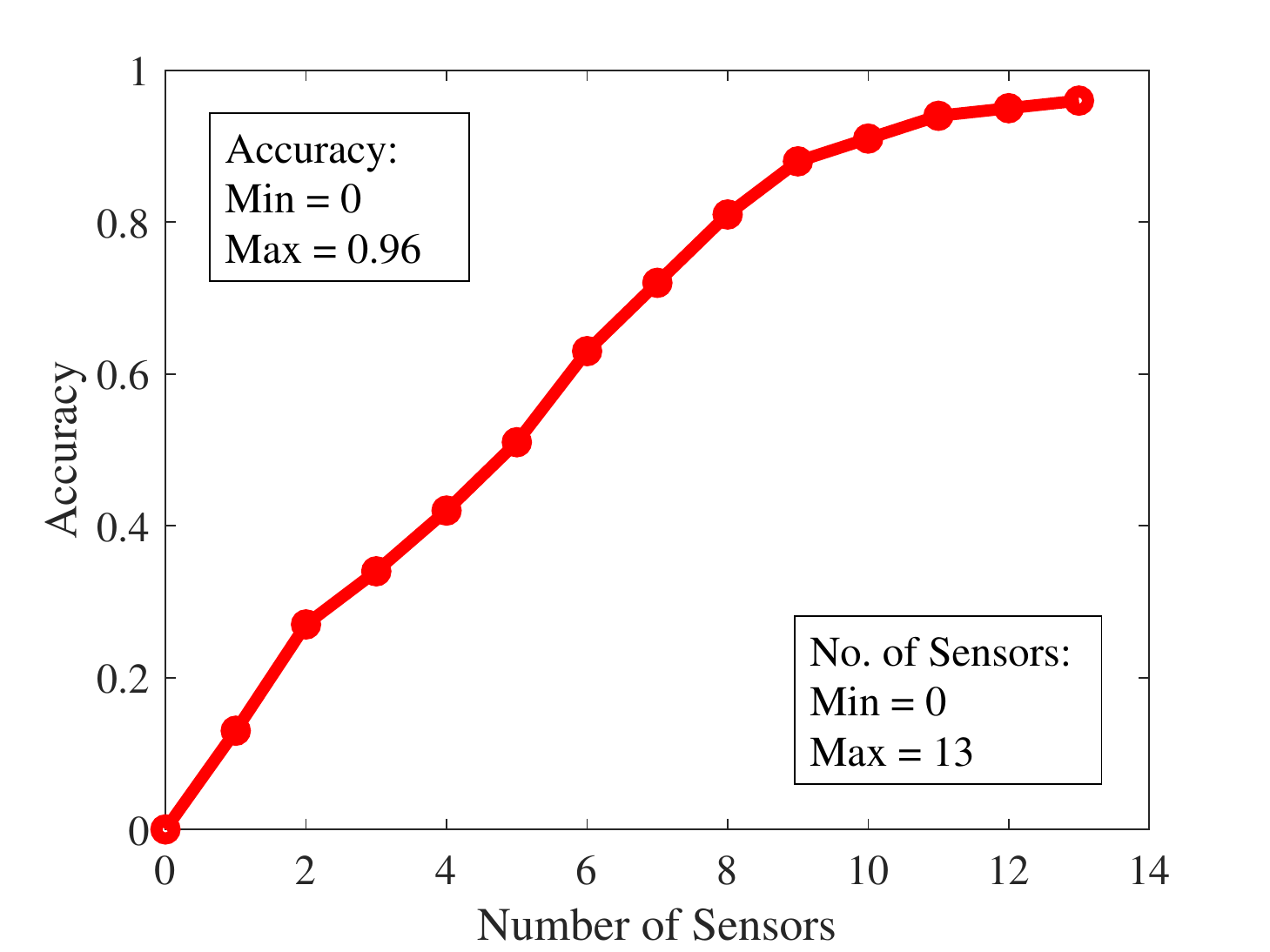}\label{fig:f12}}
\subfloat[Double Bedroom Layout]{\includegraphics[width=0.20\textwidth]{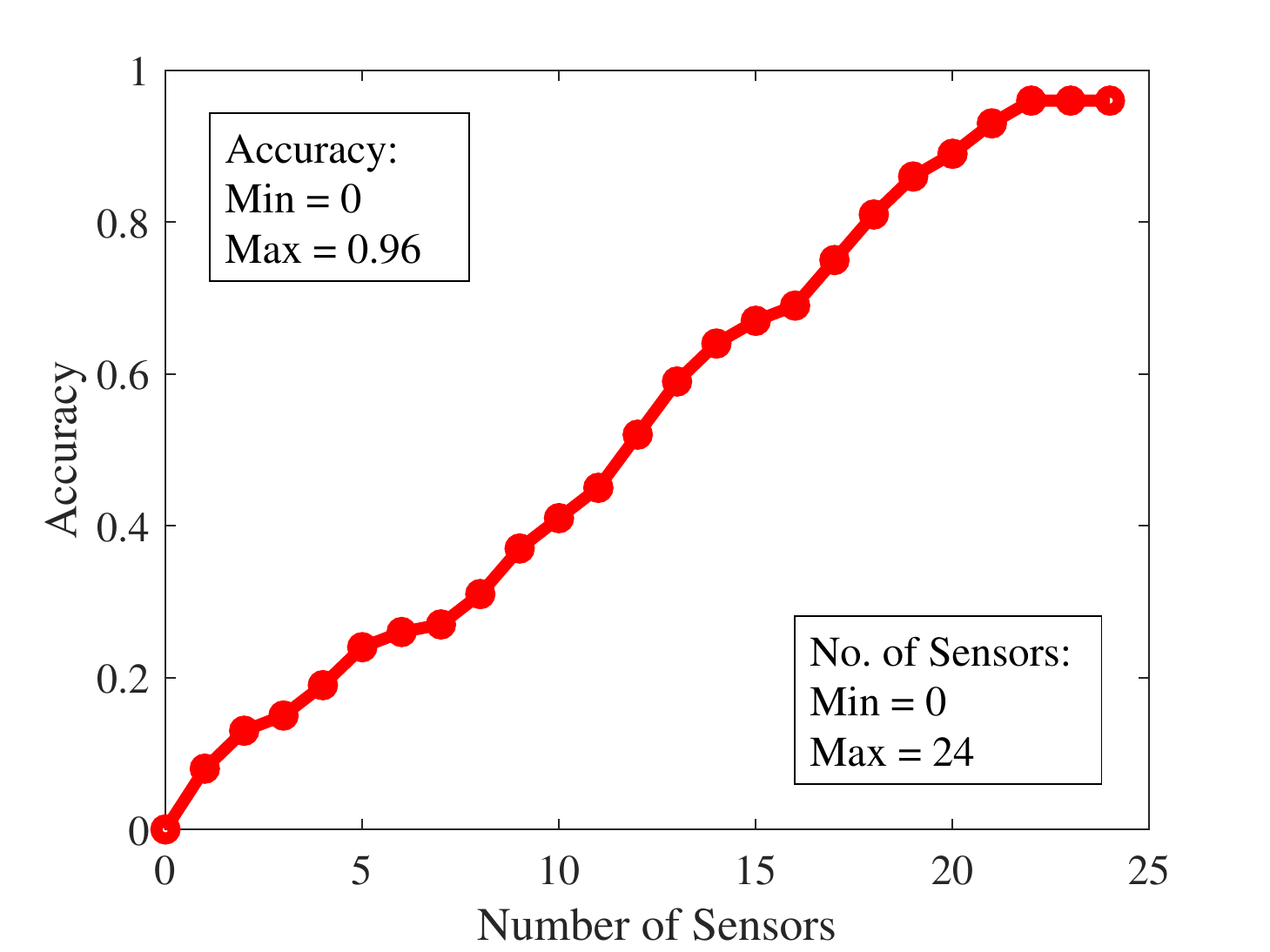}\label{fig:f13}}
\subfloat[Duplex Home Layout]{\includegraphics[width=0.20\textwidth]{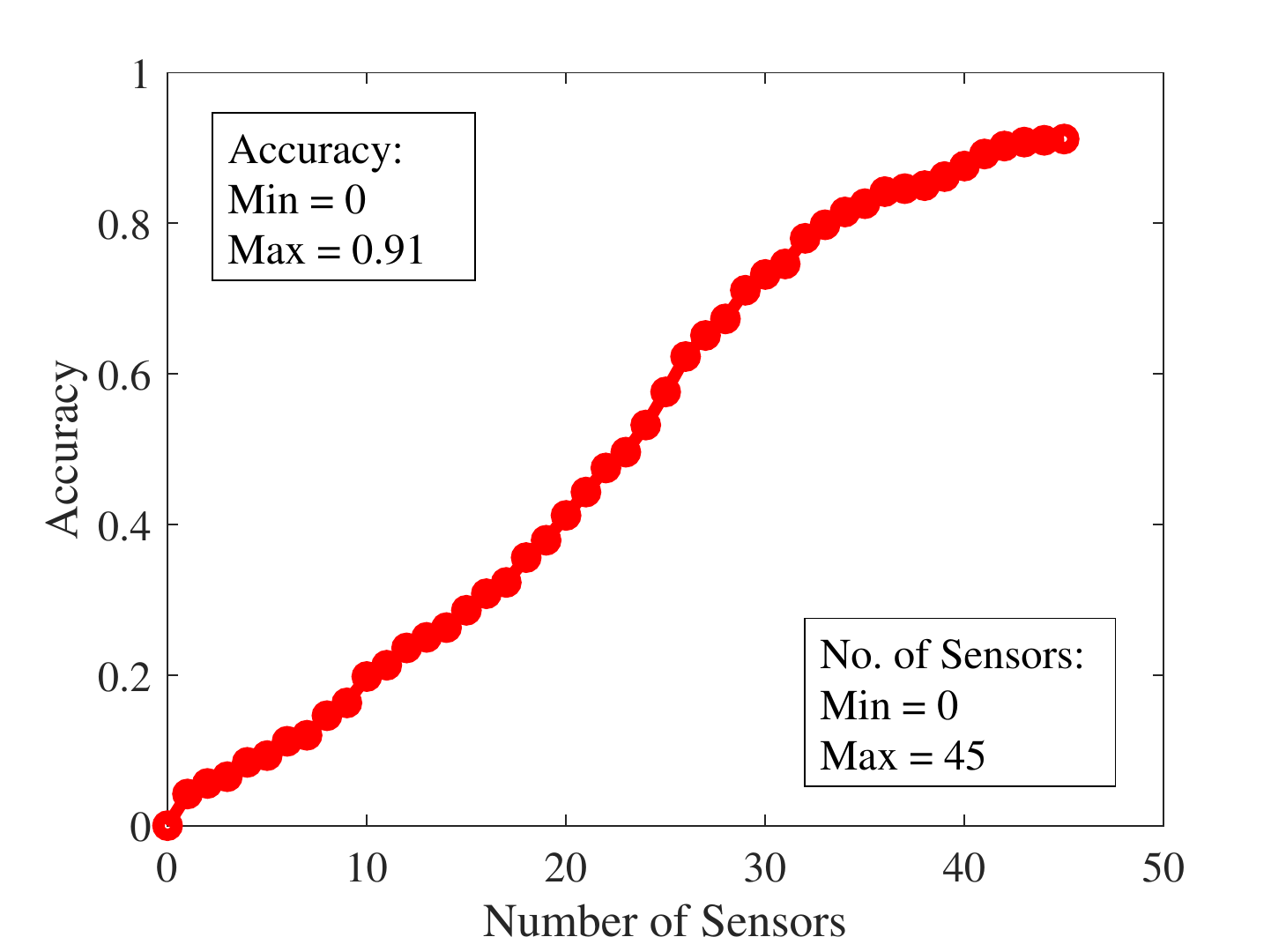}\label{fig:f14}}
\subfloat[Accuracy vs. benign apps]{\includegraphics[width=0.20\textwidth]{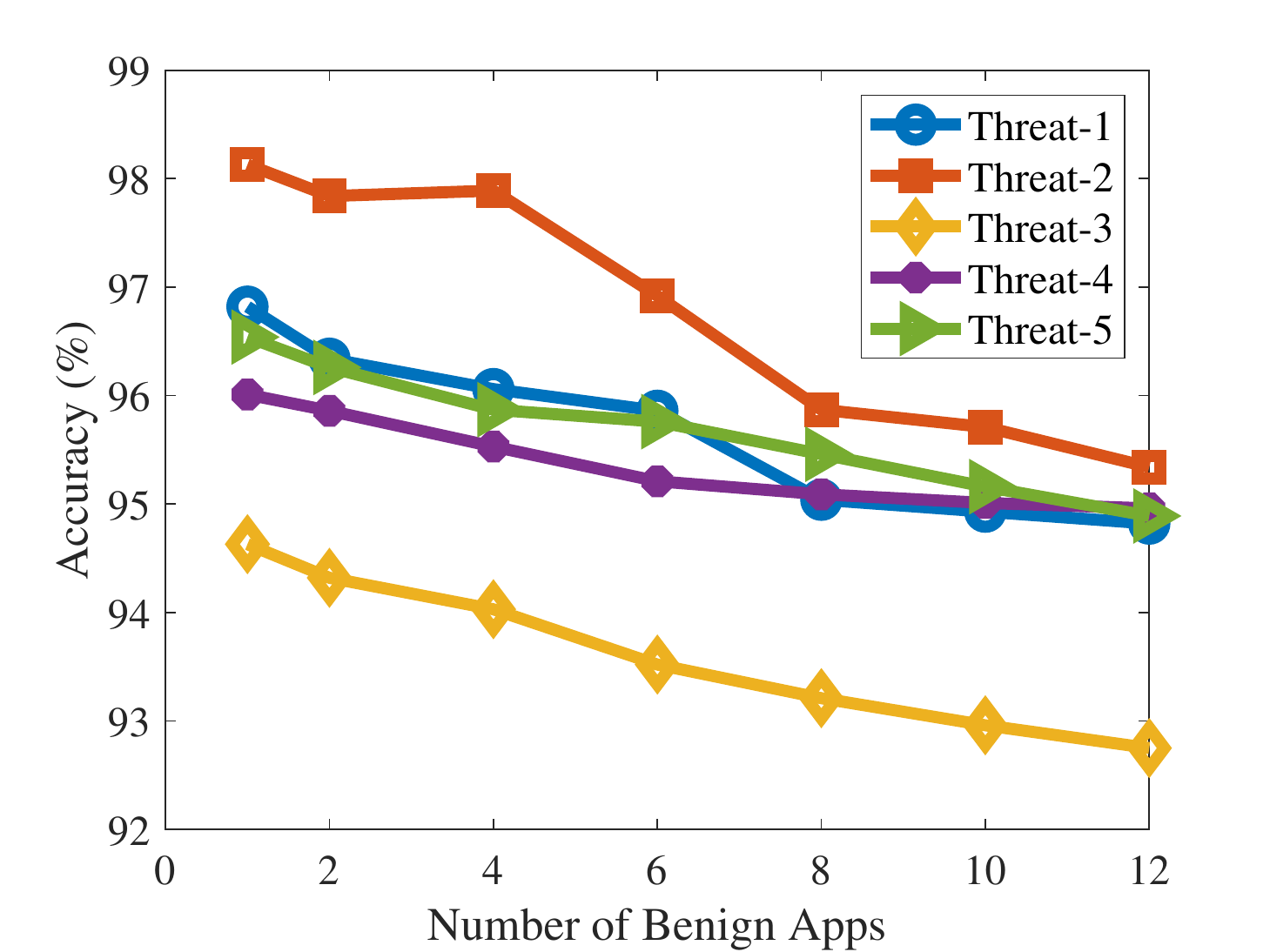}\label{figureapps}}
\subfloat[Accuracy vs. malicious apps]{\includegraphics[width=0.20\textwidth]{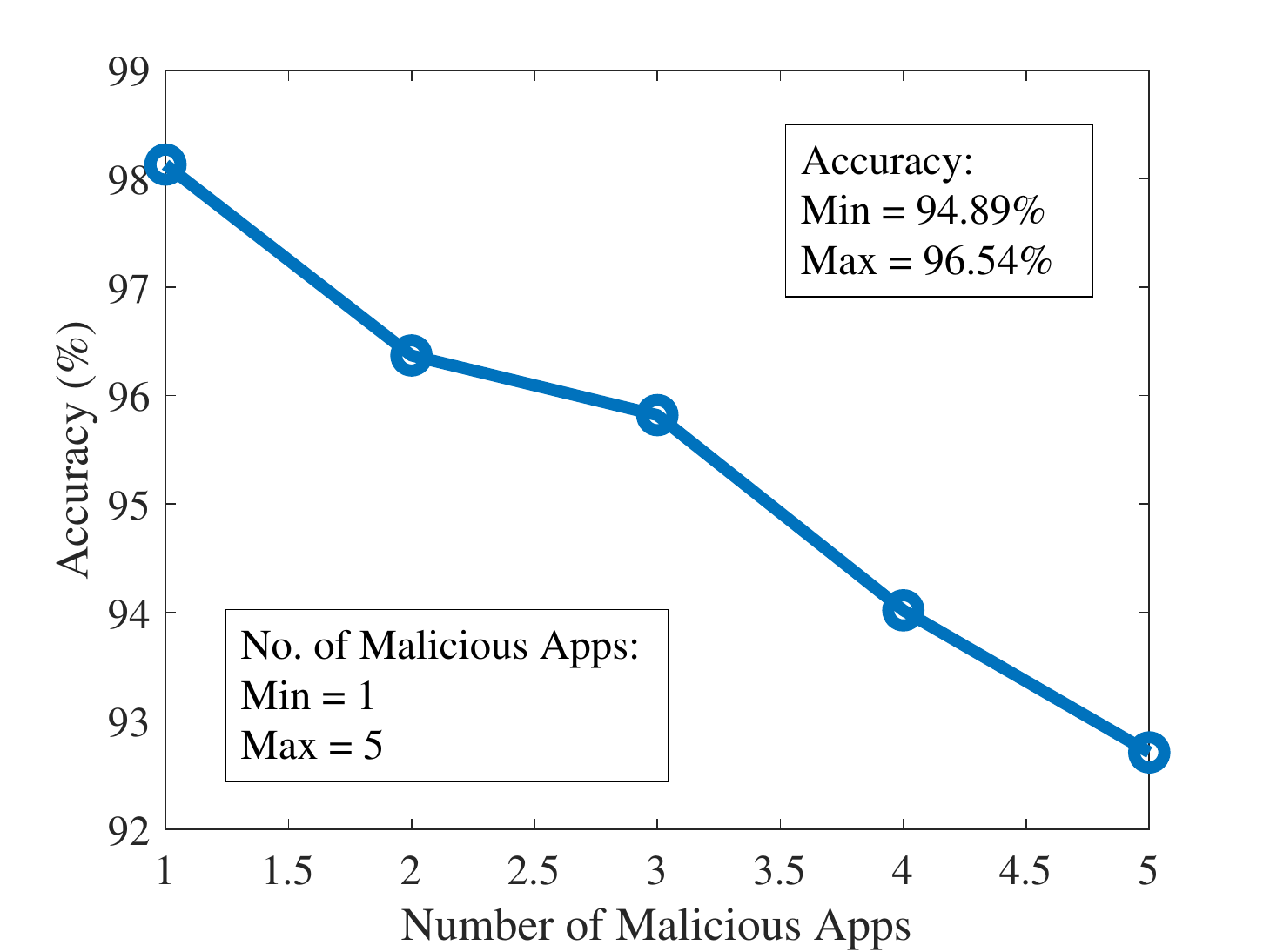}\label{figuremalapps}}
\vspace{-0.1in}
  \caption{Accuracy of \Aegis with different number of sensors (a), (b), (c) and with different number of benign and malicious apps (d), (e).}
      \label{sensors}
      \vspace{-0.3cm}
\end{figure*}

\vspace{-0.4cm}
\subsection{Evaluation with Different Smart Home Configurations}\label{configs}
In this sub-section, we evaluate \Aegis based on different smart home configurations including (1) different sensor configurations, (2) different user policies, and (3) number of installed apps.\par

\noindent\textbf{Different sensor configurations:} To evaluate the efficiency of \Aegis based on deployed sensors, we use several combinations of sensors to build the context-aware model of user activities and report accuracy in Figure~\ref{sensor}. Since \Aegis considers different smart home sensors and devices as co-dependent components, we try to understand to what extent changing the combinations of sensors in a SHS affects \Aegis's performance. For this, we tested the efficacy of \Aegis with four different combinations of sensors: without motion sensor, without the door sensor, without the temperature sensor, and without the light sensor. As seen in Figure~\ref{fig:acc} and~\ref{fig:fs}, decreasing the number of sensors from the context-aware model in \Aegis declines the accuracy and F-score of the framework. Removing the motion sensor resulted in the lowest accuracy and F-score (61\% and 68\% in duplex home layout, respectively). As motion sensors are configured with the majority of the devices (smart light, smart lock, etc.) and used in most of the user activity context, it affects the performance of \Aegis significantly. We can also observe that removing sensors from the SHS introduces high FN rate as our proposed framework cannot build the context of the user activities correctly (Figure~\ref{fig:fn}). Again, Figure~\ref{fig:acc} illustrates that removing the temperature sensor from the SHS does not influence the performance significantly (85-91\% accuracy and 88-91\% F-score in different layouts). The main reason is that the temperature sensor can be configured with a limited number of devices; hence, it is affected by user activities less than other sensors. Without the door sensor and light sensor, \Aegis can achieve moderate accuracy ranges from 77\%-86\% and 79\%-88\%, respectively. Figure~\ref{sensors} illustrates the change in accuracy of \Aegis for changing the number of sensors in different smart home layouts. For all three smart home layouts (single bedroom, double bedroom, and duplex home), limiting the number of sensors in the system decreases the accuracy of \Aegis. In conclusion, limiting the number of sensors in a SHS can reduce the efficiency of \Aegis by introducing FN cases in the system. \par

\noindent\textbf{Evaluation Based on Installed Apps:} Smart home users can install multiple smart apps to configure and control the same devices or different devices at the same time. For example, users can install two different apps to control a smart light at a time with motion and door sensor respectively. To test the effectiveness of \Aegis based on the installed apps, we installed 12 benign apps in total in the system to build the context-aware model of user activities. Figure~\ref{figureapps} shows the accuracy of \Aegis in detecting malicious apps in a SHS based on installed apps. Here, we installed different malicious apps (Section~\ref{adv}) in the system with multiple benign apps to determine the effectiveness of \Aegis. From~\ref{figureapps}, one can notice that \Aegis achieves the highest accuracy of 98.15\% for Threat-2 and the lowest accuracy of 94.34\% for Threat-3 for only one benign smart app installed in the system. With the increment of benign apps in the SHS (highest 12 benign apps), accuracy ranges between 98\% to 95\% and 94\% to 92.5\% for Threat-2 and Threat-3, respectively. The accuracy of \Aegis in detecting Threat-1, Threat-2, and Threat-5 varies between 96\% to 93\%. We also tested different malicious apps installed at once in the SHS with a fixed number of benign apps (12 benign apps) to understand the effectiveness of \Aegis completely. Figure~\ref{figuremalapps} depicts the accuracy of \Aegis based on the number of malicious apps installed on the system. One can notice that \Aegis can achieve an accuracy of 98\% for one malicious app installed in the SHS which decreases very little with the higher number of malicious apps (92.57\% with five malicious apps). In conclusion, the performance of \Aegis changed very little with the change in the number of benign apps and malicious apps installed in the SHS.

\begin{figure*}[t!]
\vspace{-0.2in}
  \centering
  \subfloat[User Policy 1]{\includegraphics[width=0.18\textwidth]{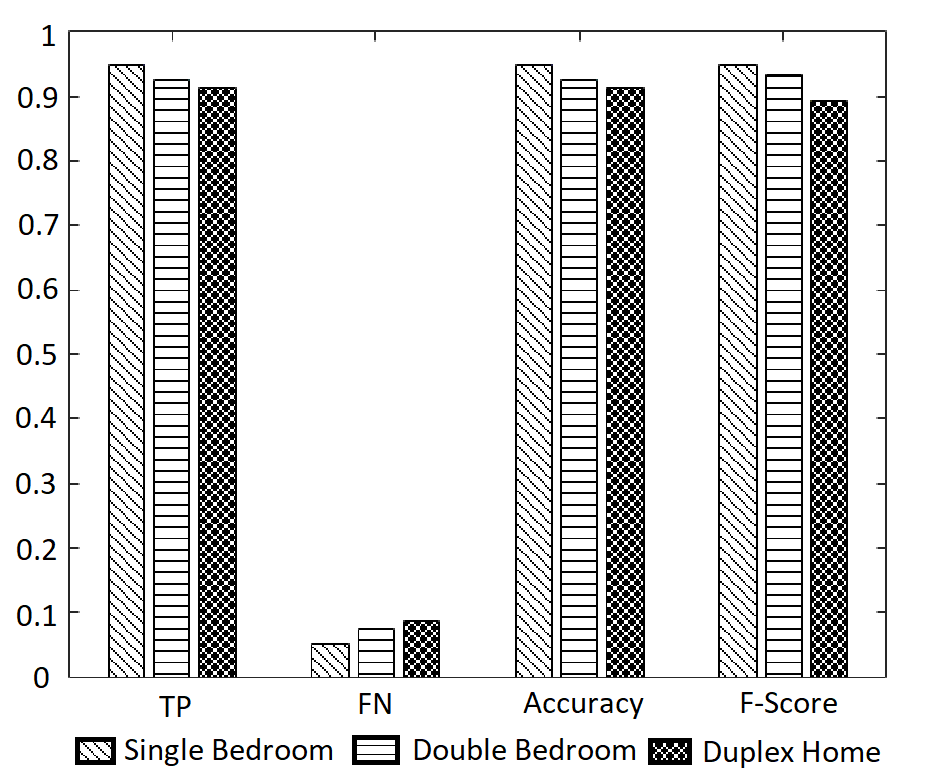}\label{fig:f8}}
\subfloat[User Policy 2]{\includegraphics[width=0.18\textwidth]{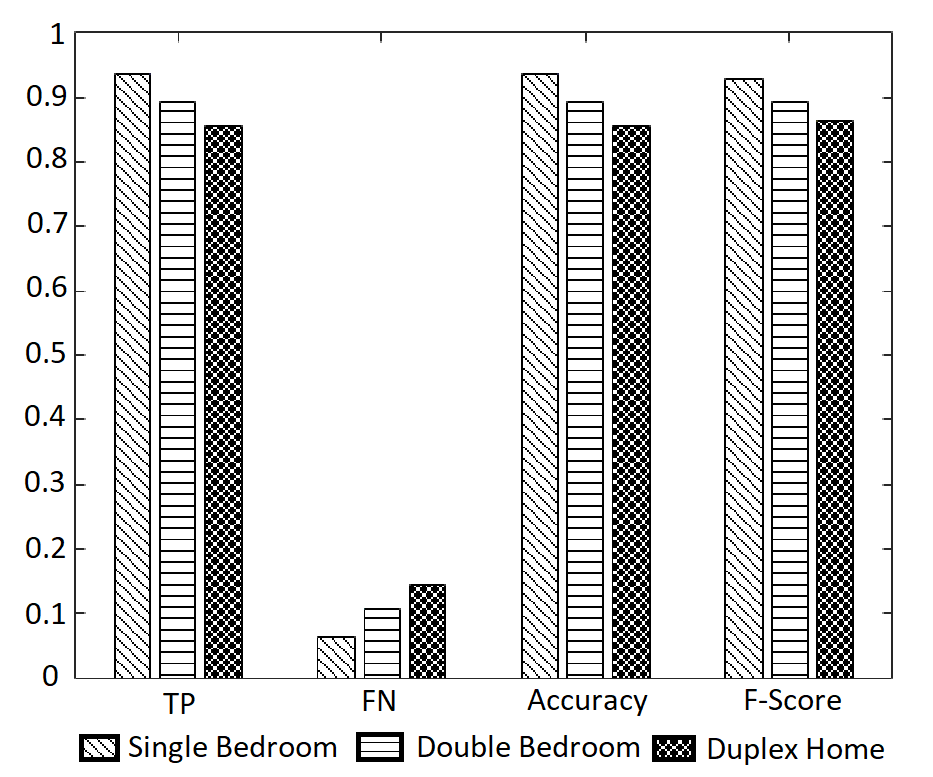}\label{fig:f9}}
\subfloat[User notification vs time]{\includegraphics[width=0.20\textwidth]{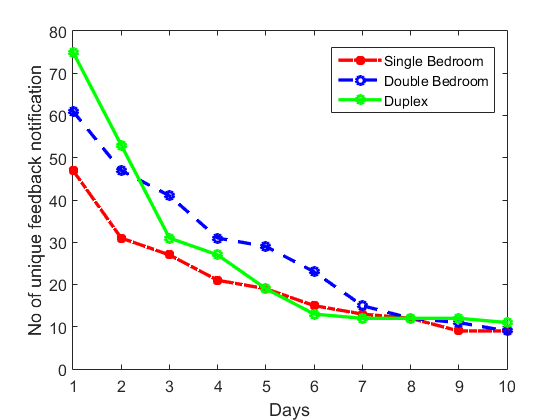}\label{FP}}
\subfloat[Accuracy vs. Feedback ]{\includegraphics[width=0.20\textwidth]{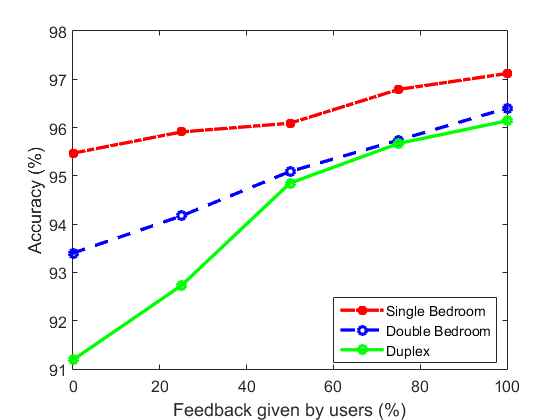}\label{acc}}
\subfloat[Training time vs. Device]{\includegraphics[width=0.20\textwidth]{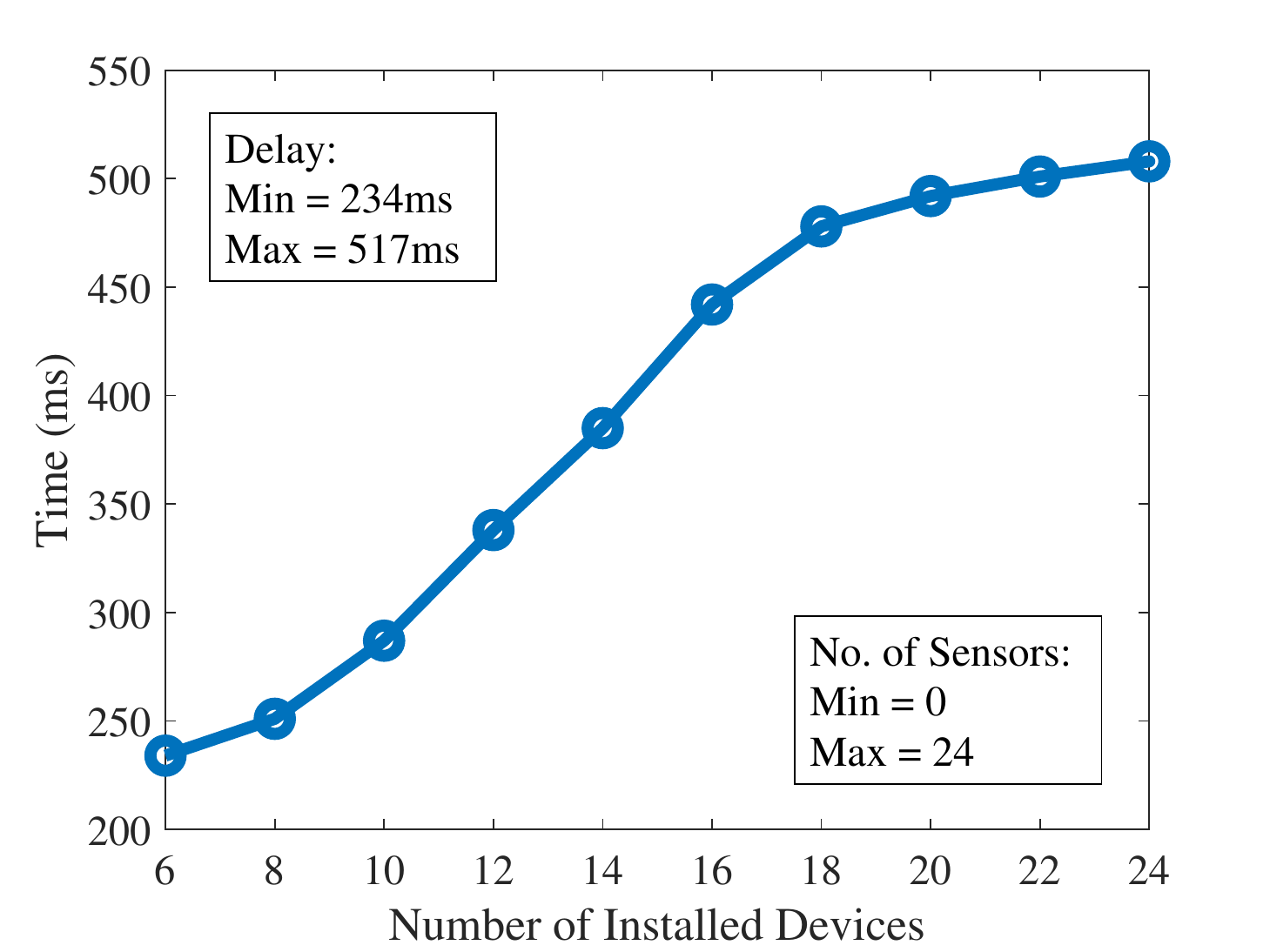}\label{adaptive}}
\vspace{-0.2cm}
      \caption{Performance evaluation of \Aegis in a policy-enforced SHS (a), (b)  and performance of \Aegis in terms of user feedback (c), (d), (e).}
      \vspace{-0.5cm}
\end{figure*}

\noindent\textbf{Evaluation Based on User Policies:} In modern SHS, users can define customized policies to control the smart home devices. For example, users can impose a time window to activate a smart light in a SHS. In this sub-section, we test the efficiency of \Aegis with different policies enforced in SHS. We consider the following user policies to evaluate \Aegis:

\noindent\textit{User Policy 1:} Users can apply time-specific operations for different smart home entities. In policy 1, users configure a smart light with the motion sensor which will be enforced only from sunset to sunrise.

\noindent\textit{User Policy 2:} Users can apply sensor-specific operations for different smart home devices. In policy 2, users configure smart lights with light, motion, and door sensors.

Figures \ref{fig:f8} and ~\ref{fig:f9} present the performance of \Aegis in these policies enforced in SHSs. One can observe that \Aegis achieved accuracy as high as 95\% while enforcing time-specific operations in SHS (Figure~\ref{fig:f8}). The F-score also ranges from 89\% to 94\% for different smart home layouts with time-specific operations with low FN rate (5\%-8\%). For User Policy 2, one can observe a slight fall in the accuracy and F-score as changing sensor-device configuration introduces FN cases in the system. From Figure~\ref{fig:f8}, we can see that \Aegis can perform with an accuracy ranging from 85\% to 93\% for different smart home layouts while changing the sensor-device configurations. \Aegis also achieves F-score ranging from 86.5-92\%  for different configurations. In summary, \Aegis can detect malicious activities in policy-enforced SHS with high accuracy and F-score. \par

\vspace{-0.1in}
\subsection{Evaluation with Different User Behavior}\label{behavior}

In this sub-section, we test \Aegis in terms of user interactions and behavior in the SHS. \Aegis uses an adaptive training method which requires users' feedback to detect FP and FN cases. This adaptive training method may cause user fatigue with excessive feedback notifications~\cite{180371}. To determine how the user fatigue may affect the performance of \Aegis, we performed accuracy vs. user feedback study which is shown in Figure~\ref{FP} and~\ref{acc}. Figure~\ref{FP} shows the number of notifications generated in adaptive training mode by \Aegis in different smart home settings over a 10 day period. One can notice that in all three layouts the number of generated notification decreases significantly. For the single home layout, the number of notifications decreases by 59\% in 5 days. For double bedroom and duplex layout, the number of notifications also decreases by 52.45\% and 74.67\%, respectively. This indicates that users only have to deal with higher feedback requests for a short period of time. Note that \Aegis pushes a notification for both FP and TN events as our test dataset includes both normal user activity and malicious events. Hence, the number of notifications generated for only FP events is lower than it seems in Figure~\ref{FP}. For example, in day 10, the total number of notification is 10 among which 6 notifications are from FP. Figure~\ref{acc} shows how user feedback affects the accuracy of \Aegis in detecting different threats. One can notice that the accuracy of \Aegis increases very little from 50\% to 100\% user feedback. This indicates that if the users actively train \Aegis in the initial period (1-5 days) in adaptive training mode, the performance improves significantly. Again, \Aegis always provides the option to choose a specific time for adaptive training mode to the users. In conclusion, \Aegis can negate the consequence of user fatigue by terminating adaptive training mode after an initial period, which is configurable by the users.
\vspace{-0.4cm}
\subsection{Performance Overhead}\label{over}
We illustrate the performance overhead of \Aegis, including resource overhead and latency. We identify two major features that could introduce a time delay in real-time operation.

\noindent\textit{Delay in adaptive training model:} \Aegis offers adaptive training mode where any malicious event detected by \Aegis is forwarded to the user for validation. \Aegis uses this validated data to retrain the analytical model which introduces a time delay in the operation. In Figure~\ref{adaptive}, we illustrate the time needed for retraining the framework with respect to the number of devices installed in the device. One can notice that \Aegis takes approximately 230 ms to train when the system has 6 different installed in the SHS. The training time increases to 519 ms for 24 installed devices in the SHS. In short, \Aegis introduces negligible overhead in terms of time delay in adaptive training mode.

\noindent \textit{Delay in action management module:} Action management module of \Aegis alerts users in the event of malicious activity in SHS. The alert message is sent to the controller device (smartphone, tablet, etc.) in the form of notification which introduces a time delay in the action management module. We use a \textit{SmartThings} app to send notifications to controller devices of authorized users. This app communicates with the cloud server via \textit{http} protocol which is connected with the action management module (Section~\ref{action}). On average, action management takes 210 ms time to send a notification to the controller device from the moment of malicious activity detection which is low for real-world deployment. In short, we conclude that \Aegis meets the efficient demands in the action management module. Appendix \ref{alertsystem} provides details of the alert system of \Aegis in normal and adaptive training mode.

\vspace{-0.2cm}
\section{User Scenarios and Discussion}\label{discussion}

In this section, we illustrate how deploying \Aegis in a smart home can help different groups of consumers using several use scenarios and discuss different benefits of \Aegis. 

\vspace{-0.3cm}
\subsection{User Scenarios}
We illustrate three different user scenarios to understand the benefits of \Aegis among vendors, end-users, and developers.

\noindent\textbf{Vendors-} Smart home vendors can use \Aegis to detect an abnormal behavior in a customer's home. Here, a customer, \textit{Alice}, installs several smart security devices (smart lock, smart camera, smart fire alarm, etc.) and the corresponding smart apps to control them. However, one of the installed app has malicious code that injects false data when no one is at home to trigger the fire alarm (Table~\ref{adversary}- Threat- 2). As Alice does not have any idea of this malicious event, she calls the security service provider/vendor for support. In this situation, the security service provider can identify that the alarm is generated from a false data from the state model generated by \Aegis and support the customer with appropriate suggestions such as deleting the malicious app, reinstalling the correct app, etc. 

\noindent\textbf{End-users-} End-users constitute the most common victims of malicious events in a smart home. Attackers can perform several malicious activities including gaining physical access to the home. For instance, a smart home user, \textit{Bob}, installs a new smart lock and the corresponding app in the SHS. However, the installed app has a malicious snippet to forward the unlock code to the attacker so he can unlock the smart lock by impersonating Bob (Table~\ref{adversary}- Threat 2). \Aegis can identify this event and notify the user in real-time. Moreover, Bob can change the state of the lock to unlock and prevent any physical access of the smart home.

\noindent\textbf{Developers-} Developers or tech enthusiastic users can deploy \Aegis in their SHS and specify different rules to enhance the security of the their homes. For instance, \textit{Kyle}, a smart home user, installs multiple smart lights and motion sensors in his SHS. Kyle also builds a new smart app to control the lights with motion. By using the logic extractor of \Aegis, Kyle can understand whether his app logic is correct or not. Moreover, Kyle can use the adaptive training mode to see how the overall state of the SHS changes with new devices and apps. If the action of the new smart lights contradicts the existing system, or any malicious event occurs (e.g., Table~\ref{adversary}- Threat 3), Kyle can understand the cause of the event and take necessary steps. Moreover, Kyle can understand the working conditions of smart home devices and improve his technological knowledge using \Aegis.

\vspace{-0.1in}
\subsection{Discussion}
\noindent\textbf{Deployability in Real-life System-} One of the prime features of \Aegis is easy deployability in real-life systems. \Aegis uses a simple smart app to collect device states in a SHS and build the context-aware model. The detection mechanism runs in the cloud which does not hamper the normal operation of the SHS. Users can install \Aegis similarly to any other smart app. 

\noindent\textbf{Applicability and Real-life Threats-} Security risks may arise from smart home apps performing side-channel attacks. For instance, a smart app can flash the light in a specific pattern to leak information or trigger another connected device which can be considered as a threat. While most of the existing solutions consider this threat as out of scope~\cite{saint-taint-analysis, mohsin2016iotsat}, \Aegis successfully detects such malicious behaviors. In addition, \Aegis can detect device malfunctions inside a SHS. For instance, if a smart light is configured with the motion sensor, one should expect that the light turns on due to the active motion. Other outcomes from this specific context may be categorized by \Aegis as a malfunction.  

\noindent\textbf{Multi-user activity in SHS-} In SHS, more than one user may perform different activities simultaneously. As \Aegis utilizes user activity contexts to detect malicious actions, correctly distinguishing between different user activities is key. Instead of single-context analysis, \Aegis uses a pattern of contexts to understand the user activities. Hence, \Aegis can detect simultaneous activities performed by different users and devices in a SHS. For instance, if two users are walking towards the same point from opposite directions, \Aegis observes the related contexts to identify two different motion activities.

\noindent\textbf{Time interval in device states-} Smart home devices use sensors to automate tasks. For instance, a smart light can be triggered by a motion sensor or a door sensor. Each trigger-action scenario has an effect time (time duration of a device being active). This effect time has to be correctly considered to build the context of the user activity. \Aegis mitigates this time dependency by considering the pattern of device utilization. For instance, the user sets a smart light to remain on for two minutes if a motion is detected. This case is detected by \Aegis by checking consecutive states of the overall smart home and is used to detect malicious apps or malfunctioning devices (if the motion is sensed by the sensor and it holds the state for 20s, the smart light should be also on for, at least, 20s otherwise broken or malicious). \Aegis uses these trigger-action scenarios to mitigate the effect of the time interval and builds the contextual model from device state patterns.

\noindent\textbf{Detecting rare events-} In a smart home, different autonomous events occur based on device configuration and user activities. These events may include rare events such as triggering fire alarms. As \Aegis uses daily user activities to train its analytical model, these rare events might be unaddressed and flagged as threat. To solve this, we use the app context to verify unrecognized events in \Aegis. Any alert triggered in \Aegis is verified with the app context generated from the installed app (Section 4.2). If the app context is matched with the rare event, \Aegis considers the event as benign and retrain the model automatically. Users can also check and verify rare natural events through action management module (Appendix~\ref{alertsystem}).

\vspace{-0.2cm}
\section{Related Work}\label{related}

In this section, we discuss threats to Smart Home Systems (SHSs) and the shortcomings of existing security solutions available for SHSs platforms.

\noindent\textbf{Security Vulnerabilities:} Recent works have outlined security threats to SHSs~\cite{notra2014experimental, schiefer2015smart, Denning}.These threats focus on three SHSs components: communication protocols, devices, and apps. Attackers may exploit implementation flaws in communication protocols to leak sensitive information from devices~\cite{fouladi2013honey, zigbee, min2015design} via information extracted from network packets~\cite{bugeja2016privacy, ho2016smart, acar2018peek}. Fernandes et al. reported several design flaws in SHSs~\cite{fernandes2016security}. Chi et al. showed that it is possible to exploit smart home platform by triggering malicious activities from legitimate user action~\cite{chi2018cross}. As smart home devices can be controlled by an accompanying smartphone app, the smartphone itself can also be used to implement attacks~\cite{sivaraman2016smart, fernandes2016security}. Jia et al. reported the existence of several malicious apps that can be migrated from smartphone and IoT platforms to SHSs~\cite{jia2017contexiot}. Recently, \textit{IoTBench} repository~\cite{iotbench-repo, saint-taint-analysis} revealed 19 different malicious apps for SHSs.

\vspace{2pt}

\noindent\textbf{Existing Security Solutions:} 
Researchers have introduced solutions to secure SHSs against existing vulnerabilities. 

\vspace{2pt}
\noindent\textit{Permission-based approach.} Previous studies proposed permission-based approaches to provide security in SHSs. Jia et al. introduced \textit{ContextIoT}, a context-aware permission model to restrict unauthorized device access and detect malicious activities in SHSs~\cite{jia2017contexiot}. 

\vspace{2pt} 
\noindent\textit{Policy and configuration analysis.} Several policy-based security measures were proposed to limit unauthorized access to SHSs~\cite{tian2017smartauth, sivaraman2015network, chakravorty2013privacy}. Similar to permission-based approaches, these solutions depend on user decision. Mohsin et al. presented \textit{IoTSAT}, a framework to analyze threats on SHSs using device configurations and enforced user policies~\cite{mohsin2016iotsat}.

\vspace{2pt} 
\noindent\textit{Static analysis.} Recently, static analysis of smart home apps have been proposed to detect information leakage and cross-app interference. Berkay and Babun et al. introduced a static analysis tool, \textit{SaINT}, to track sensitive information in smart home apps~\cite{saint-taint-analysis}. Chi et al. proposed a static analysis tool to extract app context from smart home apps to detect cross-app interference~\cite{chi2018cross}. 

\vspace{2pt} 
\noindent\textit{Forensic analysis.} Forensic analysis of smart home data has been proposed to identify malicious events in a SHS. Wang et al. proposed a security tool, \textit{ProvThings}, which logs run-time data from smart home apps and perform provenance tracking to detect malicious activities in a SHS~\cite{wang2018fear}. Babun et al. proposed \textit{IoTDots}, a forensic analysis tools which can detect user behavior from logged data in a SHS~\cite{babun2018iotdots}. 

\vspace{2pt}

\noindent\textit{\textbf{Differences from the existing solutions:}} The differences between \Aegis and existing solutions (although they are useful) can be articulated as follows. (1) While existing solutions focus on securing shared data and improving current user permission system~\cite{jia2017contexiot}, \Aegis detects malicious behaviors by considering user and device activity contexts in a SHS. (2) \Aegis considers both smart home configurations and installed apps to build a context-aware model and detect threats at run-time which outdo user-dependent solutions~\cite{jia2017contexiot}. (3) Additionally, no source code modification~\cite{babun2018iotdots} is needed for \Aegis to collect data from smart home devices and detect malicious activities in a SHS. (4) Unlike threat-specific existing solutions~\cite{saint-taint-analysis, wang2018fear}, \Aegis can detect five different types of threat in a SHS which makes it a more robust solution. (5) Finally, \Aegis collects data from a common access point and performs behavior analysis at run-time which reduces cost in terms of processing and overhead from other prior works~\cite{babun2018iotdots, saint-taint-analysis}. In addition, \Aegis does not store user data from smart home devices which reduces the privacy risks and concerns from prior solutions~\cite{babun2018iotdots}.
In summary, \Aegis offers a context-aware security framework which uses behavior analysis, usage patterns, and app context to detect malicious activities at run-time and ensures security against five different threats to SHS with high accuracy and minimal overhead.

\section{Conclusion}\label{conclusion}

New app-based smart home systems (SHSs) expose the smart home ecosystem to novel threats. Attackers can perform different attacks or deceive users into installing malicious apps. In this paper, we presented \Aegis, a novel context-aware security framework for smart homes that detects malicious activities by (1) observing the change in device behavior based on user activities and (2) building a contextual model to differentiate benign and malicious behavior. We evaluated \Aegis in multiple smart home settings, with real-life users, with real SHS devices (i.e., Samsung SmartThings platform), and with different day-to-day activities. Our detailed evaluation shows that \Aegis can achieve over 95\% of accuracy and F-score in different smart home settings. We also tested \Aegis against several malicious behaviors. \Aegis is highly effective in detecting threats to smart home systems regardless of the smart home layouts, the number of users, and enforced user policies. Finally, \Aegis can detect different malicious behavior and threats in SHS with minimum overhead. As future work, we will expand our framework by considering new multi-user settings and policies into the analysis.

\section{Acknowledgment}\label{sec:acknowledgment}
This work is partially supported by the US National Science Foundation (Awards: NSF-CAREER-CNS-1453647, NSF-1663051) and Florida Center for Cybersecurity’s Capacity Building Program. The views expressed are those of the authors only, not of the funding agencies.

{\bibliographystyle{ACM-Reference-Format}
\bibliography{Bibtex.bib}}

\appendix{}

\section{Performance Metrics}\label{metrics}
In the evaluation of \Aegis, we used six different performance metrics: True Positive rate (TP), False Negative rate (FN), True Negative  rate (TN),  False  Positive  rate (FP),  Accuracy, and F-score. TP rate indicates the percentage of correctly identified benign activities while TN rate refers to the percentage of correctly identified malicious activities. On the other hand, FP and FN illustrates the number of malicious activities identified as benign and the number of benign activities detected as malicious activities respectively. F-score is a indicator of accuracy of a framework which considers TP and TN as computational vector. The performance metrics are defined by the following equations:
\begin{equation}\small
{TP\ rate} = \frac{TP}{TP+FN} ,
\end{equation}
\begin{equation}\small
{FN\ rate} = \frac{FN}{TP+FN} ,
\end{equation}
\begin{equation}\small
{TN\ rate} = \frac{TN}{TN+FP} ,
\end{equation}
\begin{equation}\small
{FP\ rate} = \frac{FP}{TN+FP} ,
\end{equation}
\begin{equation}\small
{Accuracy} = \frac{TP+TN}{TP+TN+FP+FN} ,
\end{equation}
\begin{equation}\small
{F-score} = \frac{2*TP*TN}{TP+TN}. 
\end{equation}

\section{Sample Smart App for App Context}\label{appcontext}

One of the features of \Aegis is using app context to verify the device states in the SHS. To build the app context, we used similar static analysis approaches used in prior works~\cite{saint-taint-analysis, chi2018cross}. We performed a source-to-sink taint analysis similar to \cite{saint-taint-analysis} to extract the app context. Additionally, we consider the sources for smart apps proposed in~\cite{chi2018cross, jia2017contexiot}. We then built the abstract syntax tree (AST) and model a trigger-action scenario of an app. We tracked the \textit{Subscribe} method to represent the trigger and follow the conditional statement (e.g., if and switch) to reach the sink. This flow from entry point (source) to a sink is used to construct the condition of an app which is then represented into app context. We collected 150 official \textit{Samsung SmartThings} apps (available in their website) and created the app context database using this method. 
\modificationend
\begin{lstlisting}[caption=A code snippet of a sample smart app,label=listing-sourceCode]
/* This is a sample smart light app for Samsung SmartThings */
definition(
    name: "Smart Light App",
    namespace: "smartthings",
    author: "anonymous",
    description: "Turn lights on when door is open.",
    category: "Convenience",
)
preferences {
	section("When the door opens/closes...") {
		input "contact1", "capability.contactSensor", title: "Where?"
	}
	section("Turn on/off a light...") {
		input "light1", "capability.light"
	}
}
def installed() {
	subscribe(contact1, "contact", contactHandler)
}
def updated() {
	unsubscribe()
	subscribe(contact1, "contact", contactHandler)
}
def contactHandler(event) {
	if (event.value == "open") {
		light1.on()
	} else if (event.value == "closed") {
		light1.off()
	}
}
\end{lstlisting}

\begin{lstlisting}[caption= trigger-action Scenario of a sample app,label=listing-trigger]
Trigger: Contact1
Action: Switch1
Logic 1: contact1 = on, light1 = on
Logic 2: contact1 = off, light1 = off
\end{lstlisting}

\begin{lstlisting}[caption=Generated app context of a sample app,label=listing-context2]
App Context 1: contact1 = 1 , Light1 = 1
App context 2: contact1 = 0, Light1 = 0
\end{lstlisting}

\section{List of Smart Home Devices Used in \textsc{Aegis}}\label{devices}

\begin{table}[h!]
\centering
\footnotesize
\fontsize{6}{8}\selectfont
\resizebox{0.45\textwidth}{!}{
\begin{tabular}{{>{\centering\arraybackslash}m{1.5cm}  >{\centering\arraybackslash}m{2.8cm} >{\centering\arraybackslash}m{3.7cm} }}
\toprule
\textbf{\begin{tabular}[c]{@{}c@{}}Device \\ Type\end{tabular}} &  \textbf{Model} & \textbf{\begin{tabular}[c]{@{}c@{}}Description\end{tabular}} \\ \hline
\midrule
Smart Home Hub & Samsung SamrtThings Hub & \textbullet \hspace{1mm} Works as a central access point for smart home entities. \par \textbullet \hspace{1mm} Supports Wi-Fi, ZigBee,and Z-Wave. \\\hline
Smart Light & Philips Hue Light Bulb & \textbullet \hspace{1mm} Uses a separate communication bridge to connect with smart home hub. \par \textbullet \hspace{1mm} Uses ZigBee to communicate with other components in SHS.\par \textbullet \hspace{1mm} Supports up to 12 different sensors.\\\hline
Smart Lock & \begin{tabular}[c]{@{}c@{}}Yale B1L Lock with \\Z-Wave Push Button \\Deadbolt\end{tabular} & \textbullet \hspace{1mm} Uses Z-Wave to connect with other devices. \par \textbullet \hspace{1mm} Offers different pin code for different users. \par \textbullet \hspace{1mm} Provides both manual and remote access.\\\hline
Fire Alarm & \begin{tabular}[c]{@{}c@{}}First Alert 2-in-1 Z-Wave \\Smoke Detector and \\Carbon Monoxide Alarm \end{tabular} & \textbullet \hspace{1mm} Uses Z-Wave to connect with the hub \par \textbullet \hspace{1mm} Provides built-in smoke and CO sensors. \\\hline
Smart Monitoring System & \begin{tabular}[c]{@{}c@{}}Arlo by NETGEAR\\ Security System \end{tabular} & \textbullet \hspace{1mm} Uses Wi-Fi to connect with smart home hub. \par \textbullet \hspace{1mm} Offers both live monitoring and still pictures. \\\hline
Smart Thermostat & Ecobee 4 Smart Thermostat & \textbullet \hspace{1mm} Uses Wi-Fi to communicate with smart hub. \par \textbullet \hspace{1mm} Can be configured with sensors. \\\hline
Smart TV & \begin{tabular}[c]{@{}c@{}}Samsung 6 Series \\UN49MU6290F LED \\Smart TV \end{tabular} & \textbullet \hspace{1mm} Connects with smart home hub using Wi-Fi. \\\hline
\begin{tabular}[c]{@{}c@{}}Motion, Light,\\ \& temperature \\sensor \end{tabular} & Fibaro FGMS-001 Motion Sensor & \textbullet \hspace{1mm} Uses Z-Wave to connect with the hub. \par \textbullet \hspace{1mm} Can be configured with different devices simultaneously. \\\hline
Door Sensor & Samsung Multipurpose Sensor & \textbullet \hspace{1mm} Uses ZigBee protocol to connect with smart home hub.\\
\bottomrule
\end{tabular}}
\caption{List of devices used in the data collection.}
\vspace{-0.3in}
\label{tab:devices}
\end{table}

\section{Detailed Threat Model used during \textsc{Aegis} Evaluation}\label{threattable}
To evaluate \Aegis, we considered five different threat models (Section~\ref{adv}). We built five malicious apps to represent these threats. Our malicious apps cover several known threats presented by other researchers in~\cite{jia2017contexiot, iotbench}. To perform the attack described in Threat 1, we built a battery monitor App for smart locks that leaks the unlock code via SMS to the attacker. We realized the impersonation attack by unlocking the smart lock as an outsider using the leaked unlock code. For Threat 2, we built an app that injects false smoke sensor data to trigger the fire alarm in the SHS. For Threat 3, we created an app that flickered a smart light in a specific pattern while nobody was in the home. To perform the denial-of-service attack described in Threat 4, we developed an app that stopped the smart thermostat for a pre-defined value. For Threat 5, we created an app that could generate morse code using a smart light while no person was in the room and triggered a smart camera to take stealthy pictures. Our malicious apps cover several existing attacks on smart home devices presented by the researchers~\cite{jia2017contexiot, iotbench}. In Table~\ref{check}, we mapped our threat models with existing malicious apps presented by the researchers. Our threat model covers the malicious apps presented in the prior works and \Aegis can detect these threats with high accuracy.

\begin{table}[h!]
\vspace{-0.1in}
\centering
\footnotesize
\fontsize{6}{8}\selectfont
\resizebox{0.45\textwidth}{!}{
\begin{tabular}{cp{3cm}p{3cm}}
\toprule
\textbf{\begin{tabular}[c]{@{}c@{}}\Aegis Threat Model\end{tabular}} & \textbf{\begin{tabular}[c]{@{}c@{}}ContextIoT~\cite{jia2017contexiot}\end{tabular}} & \textbf{\begin{tabular}[c]{@{}c@{}}IoTBench~\cite{iotbench}\end{tabular}} \\ \hline
\midrule

Threat-1     & \textbullet \hspace{1mm}Backdoor pin code injection.\par 
               \textbullet \hspace{1mm}Lock access revocation.\par
                \textbullet \hspace{1mm}LockManager.\par
                \textbullet \hspace{1mm}App Update -- PowersOutAlert.\par
                \textbullet \hspace{1mm}Lock access revocation.
            & \textbullet \hspace{1mm}Permissions- Implicit 2   \\ 
Threat-2    & \textbullet \hspace{1mm}Fake alarm.\par
              \textbullet \hspace{1mm}Remote control -- FireAlarm.\par
              \textbullet \hspace{1mm}Remote command -- SmokeDetector.
            & ---     \\
Threat-3    & \textbullet \hspace{1mm}Leaking information.\par
              \textbullet \hspace{1mm}creating seizures using strobed light.\par
              \textbullet \hspace{1mm}IPC -- MaliciousCameraIPC \& PresenceSensor.\par
              \textbullet \hspace{1mm}MidnightCamera.
            & \textbullet \hspace{1mm}Side Channel - Side Channel 1.\par 
              \textbullet \hspace{1mm}Side Channel - Side Channel 1.\\ 
Threat-4    & \textbullet \hspace{1mm}Disabling vacation mode.\par
            \textbullet \hspace{1mm}Abusing permission.
            & ---      \\
Threat-5    & \textbullet \hspace{1mm}Surreptitious surveillance.\par
            \textbullet \hspace{1mm}Undesired unlocking.\par
            \textbullet \hspace{1mm}IPC -- MaliciousCameraIPC \& PresenceSensor.
            & ---      \\

\bottomrule
\end{tabular}}
\caption{Malicious Apps mapping of \Aegis, ContextIoT, and IoTBench}
\vspace{-0.5cm}
\label{check}
\end{table}

\section{Alert System in \textsc{Aegis}}\label{alertsystem}
\begin{figure}[h!]

\centering
\subfloat[Normal Mode]{\includegraphics[width=0.20\textwidth, height= 5cm]{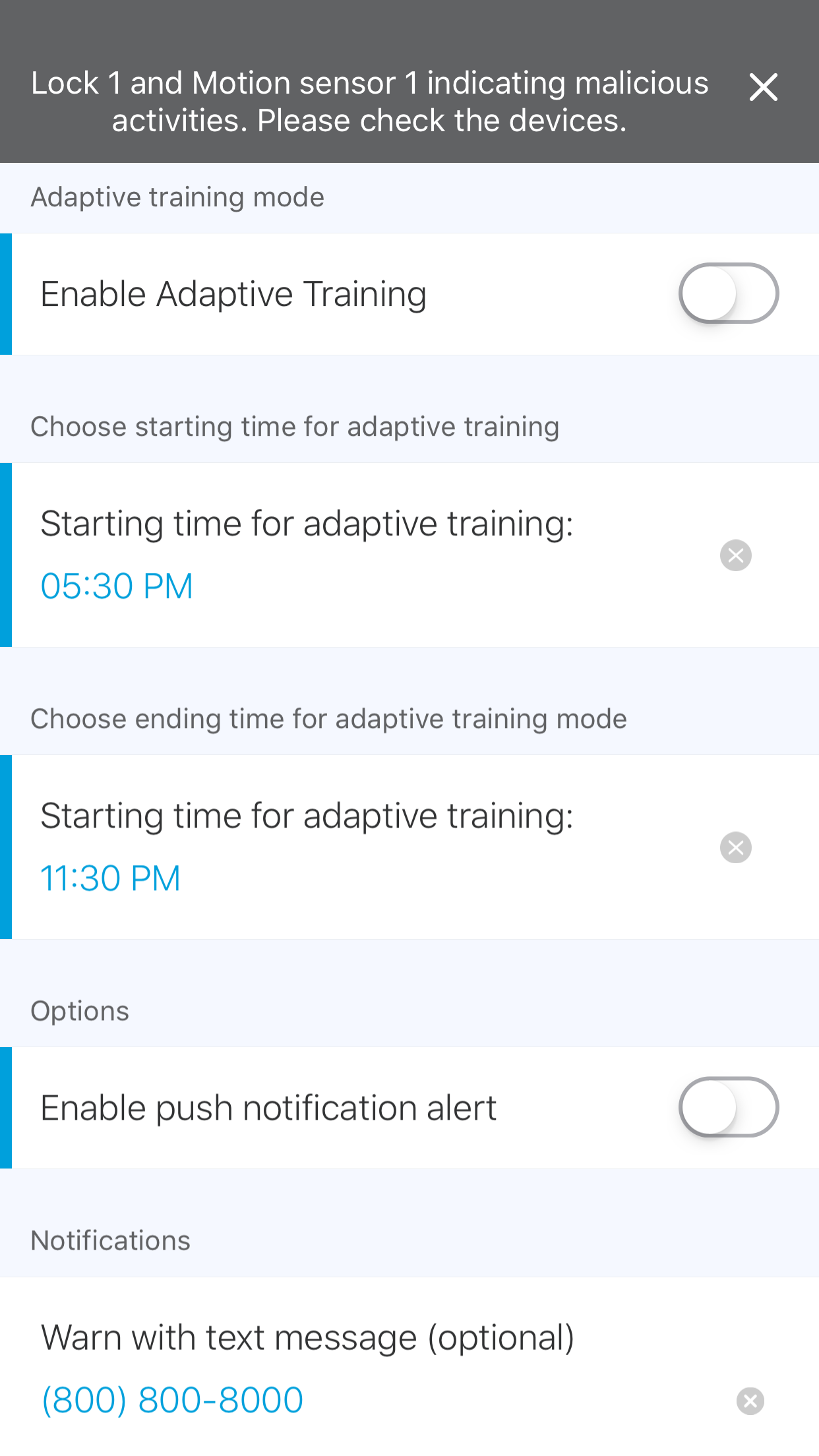}\label{fig:f10}}
\hspace{12pt}
\subfloat[Adaptive Training Mode]{\includegraphics[width=0.2\textwidth, height= 5cm]{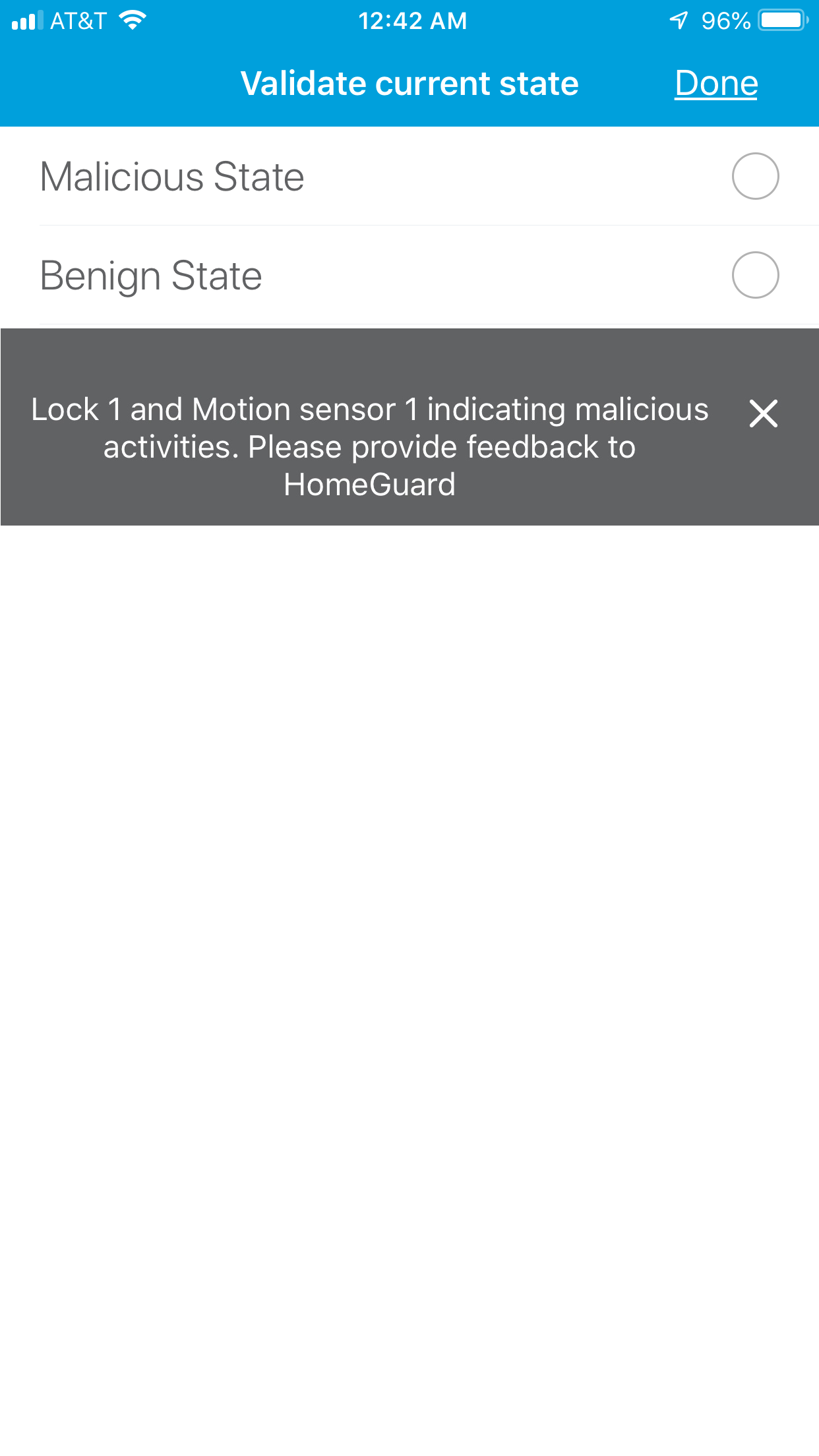}\label{fig:f11}}
\caption{User interface of \Aegis in different operation mode.}
\label{alert}

\end{figure}

\Aegis proposes an adaptive training mode to improve the accuracy of the framework and detect the false positive and negative occurrences. In Figure~\ref{alert}, both adaptive training mode and normal operation mode are illustrated.

\section{Example of SHS Layout used during Evaluation}\label{homelayout}
\begin{figure*}[h!]
\centering{\includegraphics[angle=90, width=0.6\textwidth]{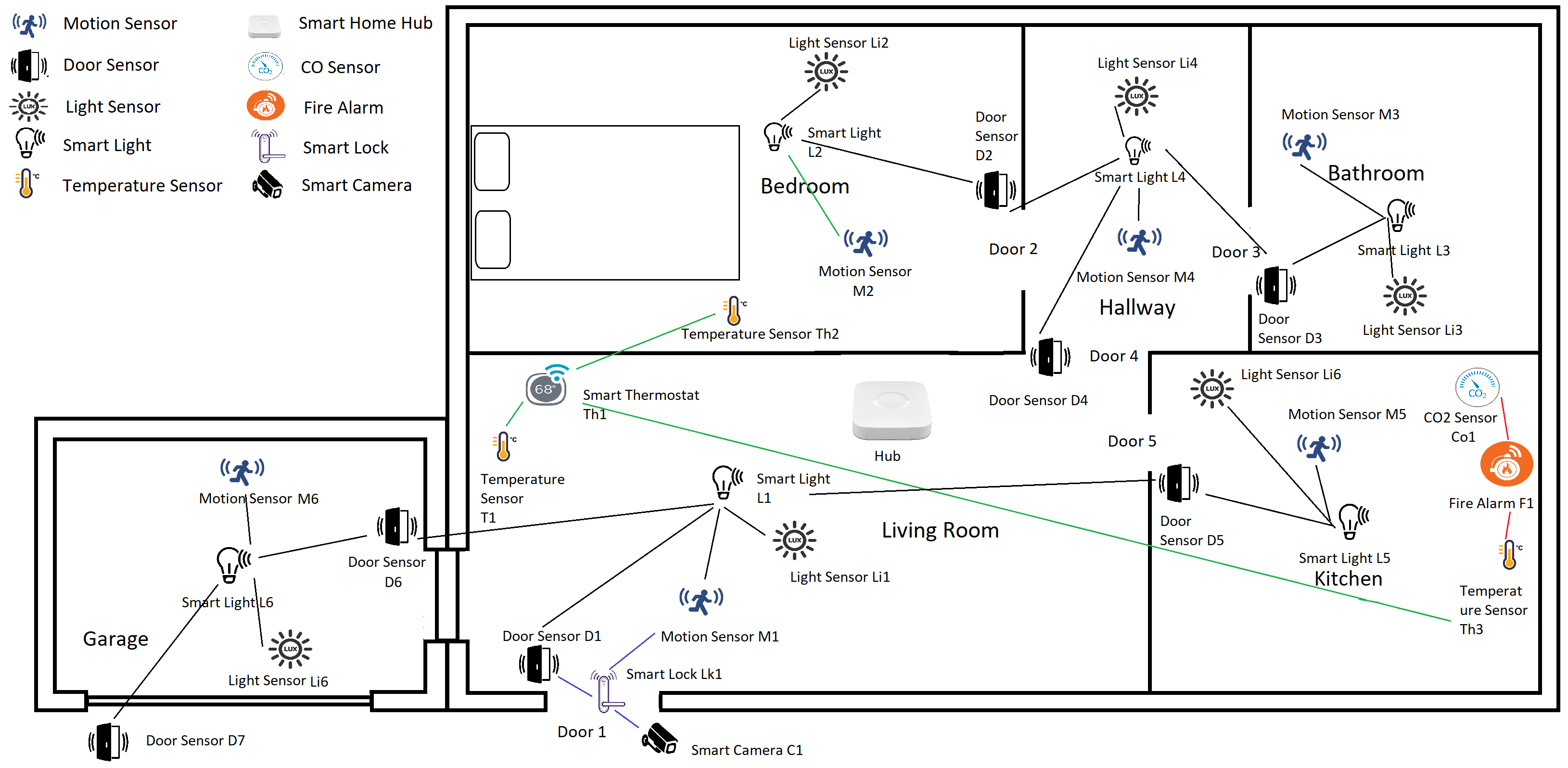}}
\caption{Single bedroom smart home layout used during evaluation.}
\label{fig:emulator}
\end{figure*}

To collect data for \Aegis, we setup three different SHS layouts where users replicated their daily activities in a smart home. In Figure \ref{fig:emulator}, the single bedroom layout of a smart home is shown which is used to take the user data in our emulator. Users clicked on different places in the layout while listing their daily activities in a timely manner. These activities were used to train and test the efficacy of \Aegis.

\end{document}